\documentclass{CUP-JNL-EDS}

\usepackage{latexsym}
\usepackage{graphicx}
\usepackage{multicol,multirow}
\usepackage{amsmath,amssymb,amsfonts}
\usepackage{mathrsfs}
\usepackage{amsthm}
\usepackage{rotating}
\usepackage{appendix}
\usepackage[numbers]{natbib}
\usepackage{ifpdf}
\usepackage[T1]{fontenc}
\usepackage{times}
\usepackage{newtxmath}
\usepackage{textcomp}%
\usepackage{xcolor}%
\usepackage{hyperref}
\usepackage{lipsum}
\usepackage{caption}
\usepackage{subcaption}
\usepackage{tikz}
\usepackage{xcolor}
\usepackage{amsmath}
\usepackage{subcaption}
\usepackage{placeins}
\usepackage{cleveref}

\definecolor{olivegreen}{rgb}{0.33, 0.42, 0.18}
\definecolor{mildskyblue}{rgb}{0.53, 0.81, 0.92}
\definecolor{turq}{HTML}{40E0D0}

\begin{document}

\begin{Frontmatter}

\title[Article Title]{End-to-end data-driven weather prediction}

\author[1]{Anna Vaughan\textsuperscript{*}\textsuperscript{\textdagger}}\email{av555@cam.ac.uk}
\author[2]{Stratis Markou\textsuperscript{*}\textsuperscript{\textdagger}}\email{em626@cam.ac.uk}
\author[2]{Will Tebbutt}
\author[3]{James Requeima}
\author[4]{Wessel P. Bruinsma}
\author[9]{\\ Tom R. Andersson\textsuperscript{\textdaggerdbl}}
\author[6]{Michael Herzog}
\author[1]{Nicholas D. Lane}
\author[8]{Matthew Chantry}
\author[5,7]{J. Scott Hosking}
\author[2,4]{Richard E. Turner\textsuperscript{*}}\email{ret26@cam.ac.uk}

\address[1]{\orgdiv{Department of Computer Science and Technology}, \orgname{University of Cambridge}, \orgaddress{\city{Cambridge}, \country{UK}}}
\address[2]{\orgdiv{Department of Engineering}, \orgname{University of Cambridge}, \orgaddress{\city{Cambridge}, \country{UK}}}
\address[3]{\orgdiv{Vector Institute}, \orgname{University of Toronto}, \orgaddress{\city{Toronto}, \country{Canada}}}
\address[4]{\orgdiv{Microsoft Research}, \orgaddress{\city{Cambridge}, \country{UK}}}
\address[5]{\orgdiv{British Antarctic Survey}, \orgaddress{\city{Cambridge}, \country{UK}}}
\address[6]{\orgdiv{Department of Geography}, \orgname{University of Cambridge}, \orgaddress{\city{Cambridge}, \country{UK}}}
\address[7]{\orgdiv{The Alan Turing Institute}, \orgaddress{\city{London}, \country{UK}}}
\address[8]{\orgdiv{European Centre for Medium-Range Weather Forecasts}, \orgaddress{\city{Reading}, \country{UK}}}
\address[9]{\orgdiv{Google Deepmind}, \orgaddress{\city{London}, \country{UK}} \\ 
\textsuperscript{\textdagger}Equal contribution authors.\\
\textsuperscript{\textdaggerdbl}Work done while at the British Antarctic Survey\\
\textsuperscript{*}Corresponding author. Email: av555@cam.ac.uk, em626@cam.ac.uk, ret26@cam.ac.uk.}

\received{10 July 2024}

\authormark{Vaughan et al.}

\abstract{

Weather forecasting is critical for a range of human activities including transportation, agriculture, industry, as well as the safety of the general public.
Machine learning models have the potential to transform the complex weather prediction pipeline, but current approaches still rely on numerical weather prediction (NWP) systems, limiting forecast speed and accuracy. 
Here we demonstrate that a machine learning model can replace the entire operational NWP pipeline. 
Aardvark Weather, an end-to-end data-driven weather prediction system, ingests raw observations and outputs global gridded forecasts and local station forecasts. 
Further, it can be optimised end-to-end to maximise performance over quantities of interest.
Global forecasts outperform an operational NWP baseline for multiple variables and lead times.
Local station forecasts are skillful up to ten days lead time and achieve comparable and often lower errors than a post-processed global NWP baseline and a state-of-the-art end-to-end forecasting system with input from human forecasters.
These forecasts are produced with a remarkably simple neural process model using just 8\% of the input data and three orders of magnitude less compute than existing NWP and hybrid AI-NWP methods.
We anticipate that Aardvark Weather will be the starting point for a new generation of end-to-end machine learning models for medium-range forecasting that will reduce computational costs by orders of magnitude and enable the rapid and cheap creation of bespoke models for users in a variety of fields, including for the developing world where state-of-the-art local models are not currently available.
}

\end{Frontmatter}

\section*{Introduction}

The numerical weather prediction (NWP) workflow, used to create the vital weather forecasts required by emergency agencies, transport providers, agriculture, energy providers and the general public builds on decades of research in earth observation, data assimilation, fluid dynamics and statistical post-processing.
From the first forecasts in April 1950, which required 24 hours to compute a single-day single-variable forecast on a 700 km grid \citep{lynch2008origins}, NWP systems have undergone a remarkable transformation.
Modern systems predict a wide range of variables at up to 15 days lead time at resolutions as fine as 10 km.
These systems consist of an intricate series of models of different components of the Earth's atmosphere, requiring bespoke supercomputers to run. \\

\begin{figure}[t!]
    \centering
    \FIG{\includegraphics[width=0.90\linewidth]{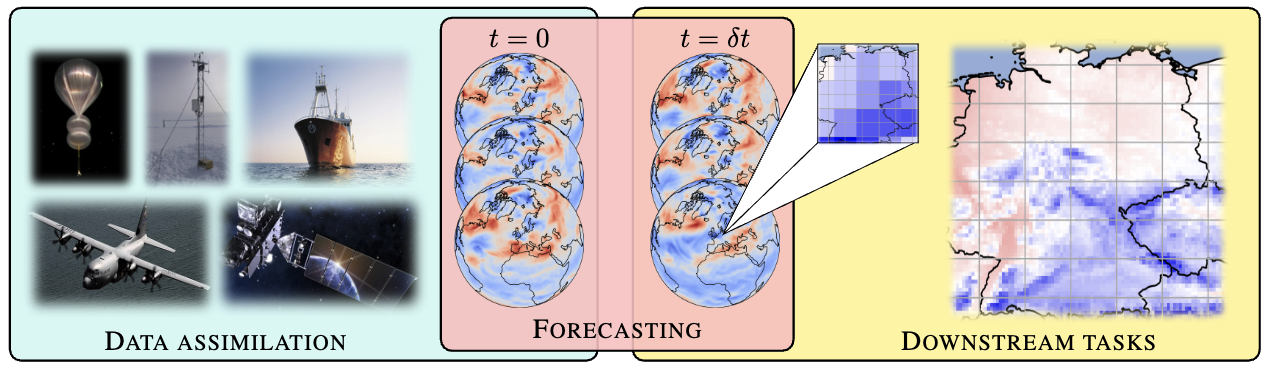}}~\\
    \vspace{-4mm}
    \FIG{\includegraphics[width=\linewidth]{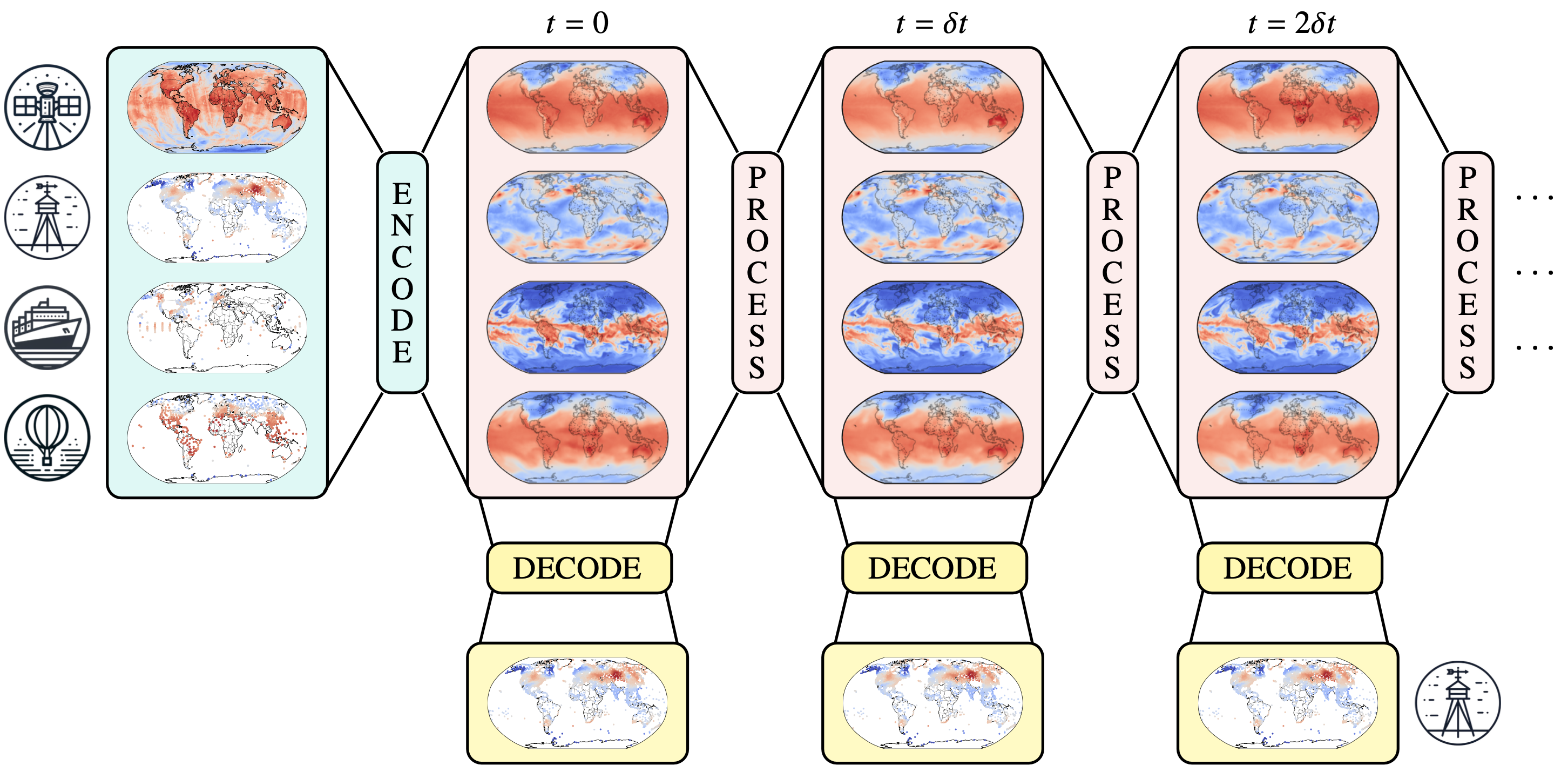}}
{\caption{
    \textbf{The weather prediction pipeline and Aardvark Weather. Top:} Illustration of the conventional end-to-end weather prediction pipeline.
    First, in the atmospheric state estimation stage (turquoise), observational data from a range of sources are used to predict the atmospheric state for multiple variables (left column of globes).
    This is used as the initial condition for the forecasting stage, which predicts the atmospheric state at future lead times (right column of globes).
    Finally, the resulting predictions are post-processed using statistical methods or further local NWP models and used for downstream applications, e.g.~generating local forecasts.
    \textbf{Bottom:} Illustration of the operation of Aardvark at deployment time.
    First, an encoder module uses raw observations as input to estimate the initial state of the atmosphere across key variables at $t = 0.$
    Next, a processor module ingests this estimated state to produce a forecast at the next lead time $t = \delta t$. Forecasts at subsequent lead times are produced autoregressively.
    Finally, a decoder module is applied to the on-the-grid states to produce off-the-grid predictions.
    The modular design of Aardvark allows for pre-training on ERA5, a large, high-quality reanalysis dataset}
\label{fig:stages}}
\end{figure}

Generating a modern weather forecast begins with the acquisition of observations from a multitude of sources, including remote sensing instruments, in-situ observations, radar systems, radiosondes and aircraft data \citep{ecmwf2023observations}.
Some of these data are processed to generate derived products such as atmospheric motion vectors and surface winds.
Raw data and resulting processed products are fed into a \emph{data assimilation} system which combines these with an initial guess from a previous forecast to generate a global approximation of the current state of the atmosphere.
This approximation is then used as an initial state for a \emph{forecasting} system that integrates equations of fluid mechanics and thermodynamics to output predictions at future lead times.
Finally, the resulting predictions from the forecasting system are used for \emph{downstream  tasks}, for example to generate local forecasts. This step may consist of statistical post-processing and running further higher resolution regional NWP models. 
Each stage of this pipeline (\Cref{fig:stages}; top) consists of multiple numerical models chained together,
resulting in an intricate workflow \citep{dueben2018challenges} that is challenging to iterate on and improve and requires bespoke supercomputers to run. This motivates the development of fast, lightweight, and customisable alternatives.\\


With end-to-end machine learning revolutionising multiple fields by replacing complex human-designed workflows, it has been suggested that a data-driven model may one day replace the entire NWP pipeline \citep{schultz2021can}.
This will be transformational for weather prediction, reducing computational costs, removing bias from inflexible aspects of NWP systems, and enabling fast prototyping and bespoke optimisation for specific tasks.
However, this has not been attempted to date, with studies focusing on applying machine learning to the easiest components of the pipeline.
For example, machine learning models have been shown to outperform their operational state-of-the-art counterparts for replacing the numerical solver in the forecasting component \citep{keisler2022forecasting,bi2022pangu,lam2023graphcast,price2023gencast,chen2023fengwu,chen2023fuxi,bodnar2024aurora}, deriving variables from raw satellite data in pre-processing \citep{shao2023machine,yan2020deep,zhang2018retrieval}, and post-processing forecast data in the downstream stages \citep{kirkwood2021framework,gronquist2021deep}.
Work on replacing the most challenging component, the assimilation system, remains at the stage of developing initial prototypes \citep{chen2023towards,huang2024diffda,xu2024fuxi}.
The vision of an end-to-end data-driven solution therefore remains aspirational, with conventional NWP systems essential for all forms of operational forecasting. \\  

In a recent article assessing the prospect of end-to-end deep learning weather prediction the verdict was that “a number of fundamental breakthroughs are needed before this goal comes into reach” \citep{schultz2021can}.
As late as the start of 2024 it was thought by an expert well versed in the AI revolution in weather forecasting that full end-to-end systems are “probably many years in the future” \citep{Rasp2024}.
Here we report that these breakthroughs are happening earlier than expected.
We present Aardvark Weather, the first end-to-end data-driven weather forecasting system capable of generating predictions with no input from conventional NWP by instead learning a mapping from raw input observations to output forecasts.
This allows Aardvark to tackle the complete weather prediction pipeline whilst being entirely independent from NWP products at prediction time, relying solely on observation data to generate forecasts.
We demonstrate that, using an order of magnitude fewer observations than those available to operational baselines and orders of magnitude less compute, Aardvark is capable of producing forecasts on a global $1.41^{\circ}$ grid that achieve lower root mean squared error (RMSE) than operational NWP systems across multiple variables and lead times. 
Furthermore, we demonstrate that this system provides local forecasts that achieve lower errors than post-processed NWP and a full end-to-end operational forecasting system for multiple lead times, and can be optimised end-to-end to maximise performance over variables and regions of interest. \\

\section*{Results}

\subsection*{Aardvark Weather}
\label{subsec:results:aarvark}
Aardvark Weather is a deep learning model which provides forecasts of eastward wind, northward wind, specific humidity, geopotential and temperature at 200, 500, 700 and 850hPa pressure levels, 10-metre eastward wind, 10-metre northward wind, 2-metre temperature and mean sea level pressure on a dense global grid, as well as station forecasts for 2-metre temperature and 10-metre wind speed.
Aardvark consists of three modules, and is designed to leverage high-quality reanalysis data during training while being entirely independent from NWP products at deployment time.
\Cref{fig:stages} (bottom) illustrates the operation of Aardvark at deployment, outlining the function of each of its three modules.
First, an \emph{encoder} module takes in observational data from a multitude of sources, both on-the-grid and off-the-grid, and produces a gridded initial state.
To achieve this we leverage recent advances from deep learning \cite{garnelo2018conditional,gordon2019convolutional} in handling off-the-grid and missing data.
We note that our approach for state estimation differs from the data assimilation systems used in conventional NWP pipelines.
In data assimilation the initial state is estimated using a forecast from the previous run as a first guess, and updating this using incoming observations to estimate the current atmospheric state.
In contrast, we take inspiration from the success of machine learning methods in end-to-end systems, and map directly from observational data to forecasts without relying on a recurrent structure.
This provides a system that is simple to train and deploy, and does not suffer from instabilities and degradation over time.
Once the initial atmospheric state has been estimated, it is used as input to a \emph{processor} module, which produces a gridded forecast at a lead time of 24 hours.
Forecasts at subsequent lead times are produced by autoregressively feeding the predictions of the processor module back to it as an input, similar to existing approaches in data-driven weather forecasting \citep{bi2022pangu,lam2023graphcast,nguyen2023scaling}.
Finally, task-specific \emph{decoder} modules ingest these forecasts and produce local predictions.
While in this work we consider a decoder for a single downstream task, producing local station forecasts, this system is suitable for use with multiple separate decoders for different tasks.
Together, the encoder, processor and decoder modules form a neural process \citep{garnelo2018conditional,gordon2019convolutional}, a machine learning system which naturally handles off-the-grid and missing data.
A vision transformer \citep[ViT;][]{dosovitskiy2020image} forms the backbone of the encoder and processor modules, while decoder modules are implemented as a lightweight convolutional architecture. \\

\begin{figure}[t!]
    \centering
    \FIG{\includegraphics[width=\linewidth]{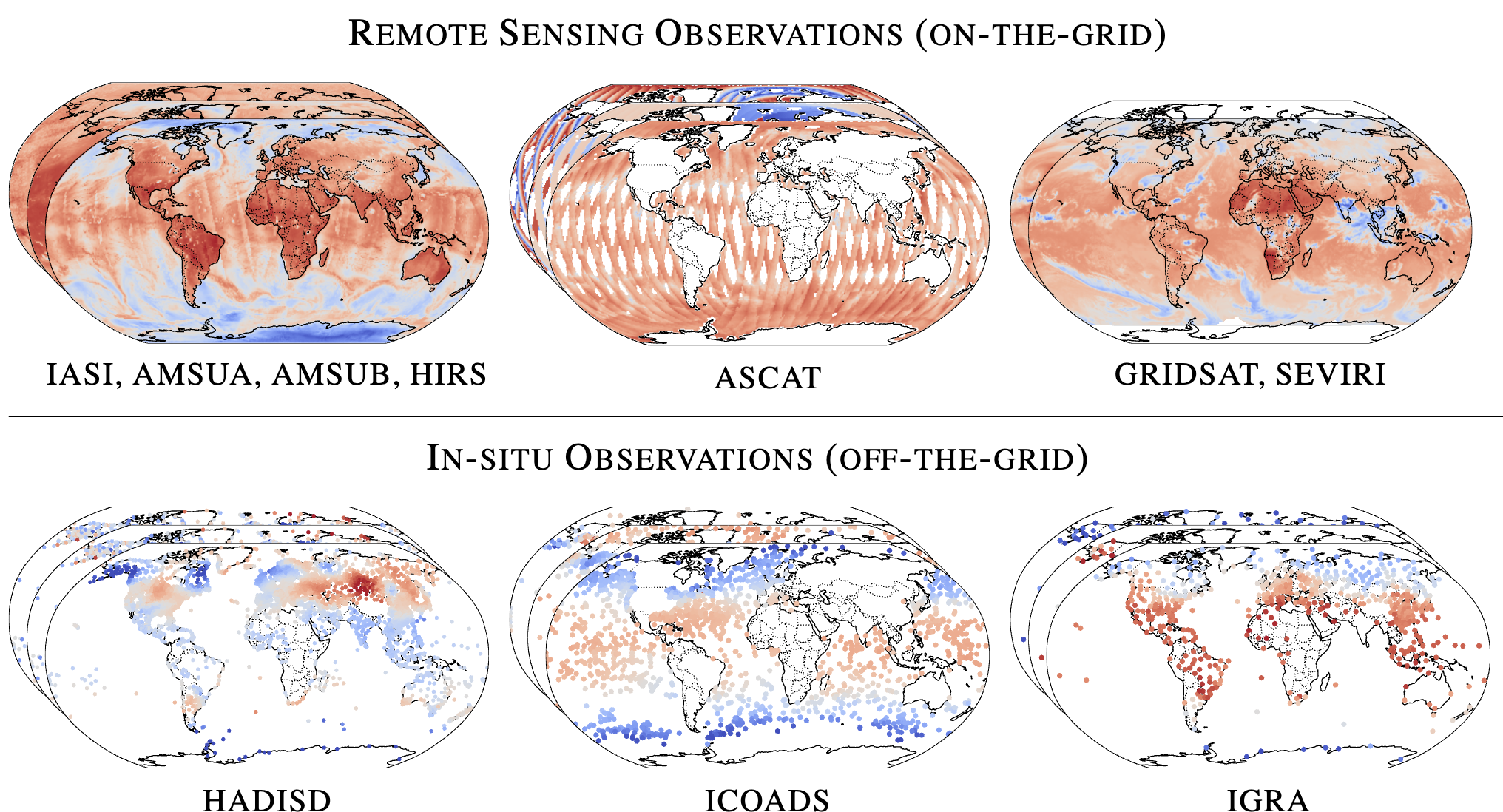}}
{\caption{
    \textbf{Illustration of the different data sources leveraged in Aardvark.}
    The input data to Aardvark consist of a combination of observations from remote sensing instruments (top row) which we pre-grid before passing to the model, as well as in-situ observations from land and marine observation platforms and radiosondes (bottom row).
    Each of these data modalities contain several observational variables, of which we select a subset here for the purposes of illustration.
    The remote sensing data also include a range of meta-data about the measurements,  omitted here for simplicity.
    White areas indicate regions of missing data which must be handled by the encoder module}
\label{fig:data}}
\vspace{7mm}
\end{figure}

A key challenge in designing machine learning systems for observational atmospheric data is that the record for many instruments is relatively short, limiting the data available for training.
The modular design of Aardvark (\cref{fig:stages}; bottom) avoids this issue by enabling pre-training using high-fidelity historical reanalysis data before fine-tuning on the scarcer observational data.
Specifically, we train the system in a way that mimics how it will be deployed.
We start by pre-training the encoder module using raw observations as input and reanalysis data as output.
We also pre-train the processor using reanalysis data for both inputs and outputs, and then fine-tune using the estimated states.
We next train the decoder using the output of the processor as input and raw data as targets.
This procedure ensures there is no mismatch between the training and deployment of the system.
Finally, we fine-tune the encoder, processor and decoder modules jointly, to optimise to the entire model for a specific variable and region. For all modules we train on data prior to 2018, and hold out 2018 and 2019 as test and validation years respectively. \\

\subsection*{Input variables}
\label{subsec:results:input}
Accurately estimating the state of the atmosphere requires inputs from a variety of observation sources.
Input variables are selected to capture dynamics both at the Earth's surface as well as at multiple different levels through the atmosphere.
In-situ observations are taken from weather stations and ships at surface level, and radiosondes at upper levels.
As coverage from these instruments is largely confined to the surface, as well as geographically skewed and sparse, remote sensing instruments provide a crucial complementary global data source.
Motivated by gains observed in operational NWP systems \citep{laloyaux2016impact,isaksen2004impact,eyre2022assimilation}, we select four primary sources of satellite data: scatterometer data to provide information about surface wind, microwave multispectral infrared and hyperspectral infrared sounder data to provide information on upper atmosphere temperature and humidity profiles, and geostationary sounder data to provide an instantaneous snapshot of the state of the atmosphere.
These observations are taken with different time windows ranging from one to 24 hours prior to lead-time zero. 
In contrast to HRES, no observations past the analysis time included \citep{ecmwf_newsletter_2020}.
\Cref{fig:data} shows an example of a single time slice of input data to Aardvark for in-situ and remote sensing sources.
These atmospheric observations are augmented by several temporal and orographic variables.
We note that Aardvark only ingests approximately 8\% of the observations available to conventional NWP systems, more than an order of magnitude less input data. \\

\subsection*{Performance evaluation: global forecasting}
\label{subsec:results:forecasting}
For global gridded forecasts we compare Aardvark to four baselines.
The simplest of these, allowing for assessment of whether a forecasting system is skillful, are persistence and hourly climatology.
The final comparison is to the two most widely used deterministic operational NWP systems: HRES from the European Centre for Medium Range Weather Forecasting and GFS from the National Centers for Environmental Prediction.
Of these two baselines HRES typically outperforms GFS on global metrics, however we include GFS because operational centres often use a selection of different models, including GFS, to create their local forecasts including the National Weather Service which we use as a baseline for local forecasting experiments.
For each variable, pressure level, and lead time, we evaluate and report latitude weighted RMSE, a common metric for assessing the performance of deterministic forecasting systems \citep{rasp2023weatherbench}.
For all baselines we take ERA5 reanalysis as ground truth. \\

Figure 3 shows latitude weighted RMSE performance compared to the baselines for eight headline variables.
For two metre temperature and mean sea level pressure Aardvark provides skillful predictions at up to nine days lead time, outperforms GFS at all lead times and approaches the performance of HRES at longer lead times.
For 10-metre eastward wind and 10-metre northward wind Aardvark provides skillful forecasts out to eight days lead time, with similar errors to GFS from zero to four days, outperforming GFS from five days onwards and HRES from seven days onwards.
For temperature at 850hPa Aardvark outperforms GFS at all lead times and HRES from day eight onwards.
For specific humidity at 700hPa Aardvark performs comparably to GFS at days zero to one and outperforms GFS from day two onwards and HRES from day five onwards.
Geopotential at 500hPa is more challenging for Aardvark, with higher errors than HRES and GFS out to eight days.
Finally, for eastward wind at 700hPa Aardvark has slightly higher errors than GFS out to four days and outperforms it thereafter, while catching up to and outperforming HRES from day seven onwards. \\ 

Figure \ref{fig:initial_condition} shows an example of gridded global predictions at lead times of zero, one, two and four days for 10-metre eastward wind.
Complete plots of predictions for all variables are given in appendix \ref{app:forecast_full}. Aardvark successfully captures large scale features of the atmospheric state, accurately predicting both mid-latitude systems and tropical features. Many details are well represented, for example the formation and track of a tropical cyclone in the Southern Indian Ocean is successfully forecast. 
This example demonstrates the potential of Aardvark for forecasting mesoscale high-impact weather events. 
Although some spectral blurring of the higher spatial frequencies is evident, these results are of remarkably high fidelity given the limited resolution and range of observational data provided to the model.  \\

There are several factors to consider when interpreting these results. Overall, Aardvark's errors are larger at higher atmospheric levels and shorter lead times compared to the operational baselines.
This is most likely due to the higher concentration of observations close to the surface, a common issue with data-driven models for the forecasting component of the pipeline. 
For longer lead times a by-product of finetuning the forecasting module to minimise errors at future lead times is that forecasts tend to become spectrally blurred, a phenomenon also commonly seen in data-driven weather forecasting systems \citep{ecmwf_fuxi_2023}.

\begin{figure*}[t!]
\centering
\includegraphics[width=\linewidth]{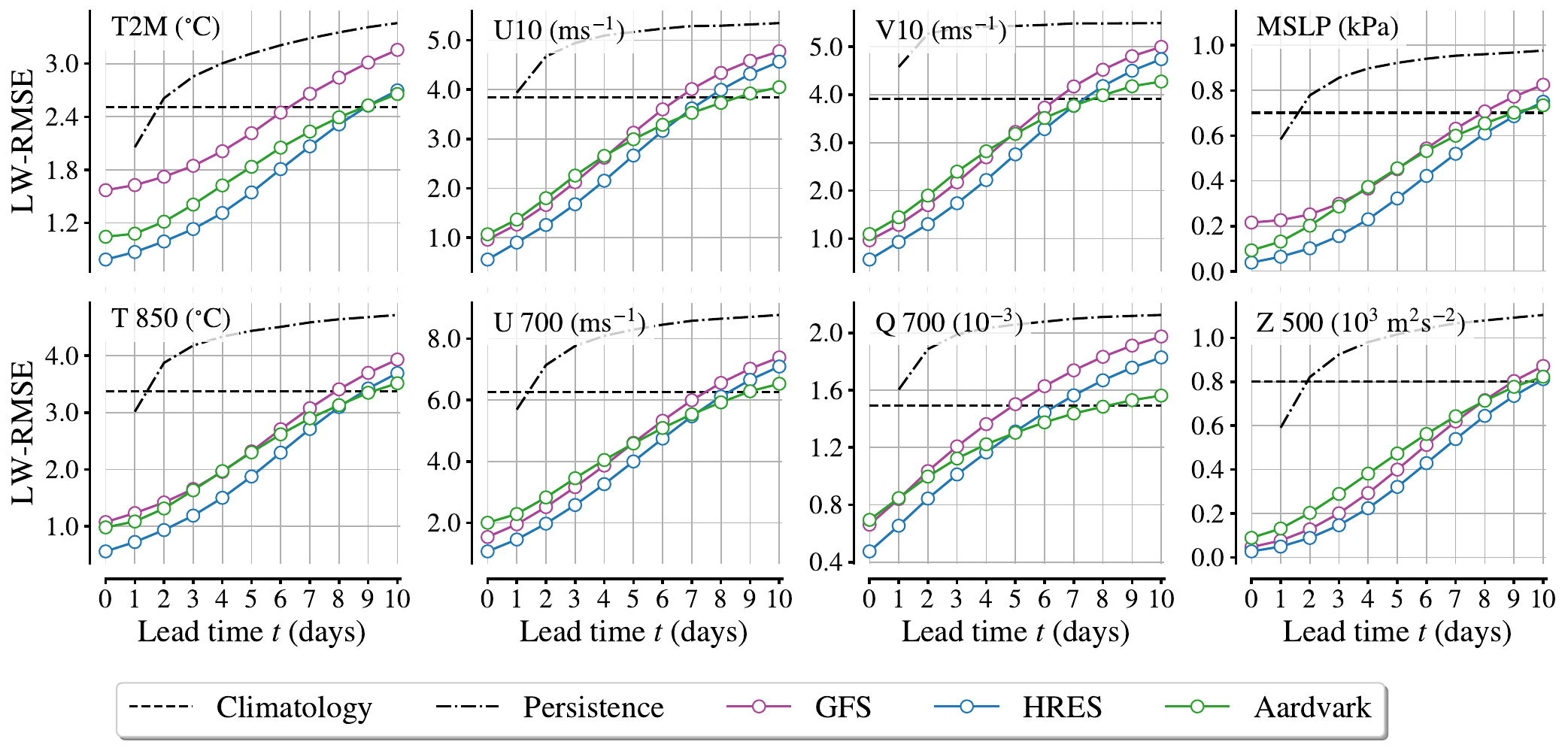}
\vspace{-7mm}
\caption{
\textbf{Gridded global forecast performance for selected variables.}
Latitude-weighted RMSE using ERA5 reanalysis data as the ground truth, on the held out test year (2018), for the four surface variables: 2-metre temperature (T2M), 10-metre eastward wind (U10), 10-metre northward wind (V10) and mean sea level pressure (MSLP) (first row) and four headline upper-atmosphere variables: temperatre at 850hPa (T850), eastward wind at 700hPa (U700) specific humidity at 700hPa (Q700) and geopotential at 500hPa (Z500) (second row), as a function of lead time $t.$
At lead time $t = 0,$ Aardvark predicts the initial atmospheric state from from observational data alone.
The error at $t = 0$ corresponds to the error in the initial state.
Note that HRES has non-zero error at $t = 0,$ as it is compared to ERA5 reanalysis ground truth.
Results for the full set of variables predicted by Aardvark are given in appendix \ref{app:forecast_full}}
\label{fig:forecast}
\vspace{-4mm}
\end{figure*}

\begin{figure}
\centering
\includegraphics[width=\linewidth]{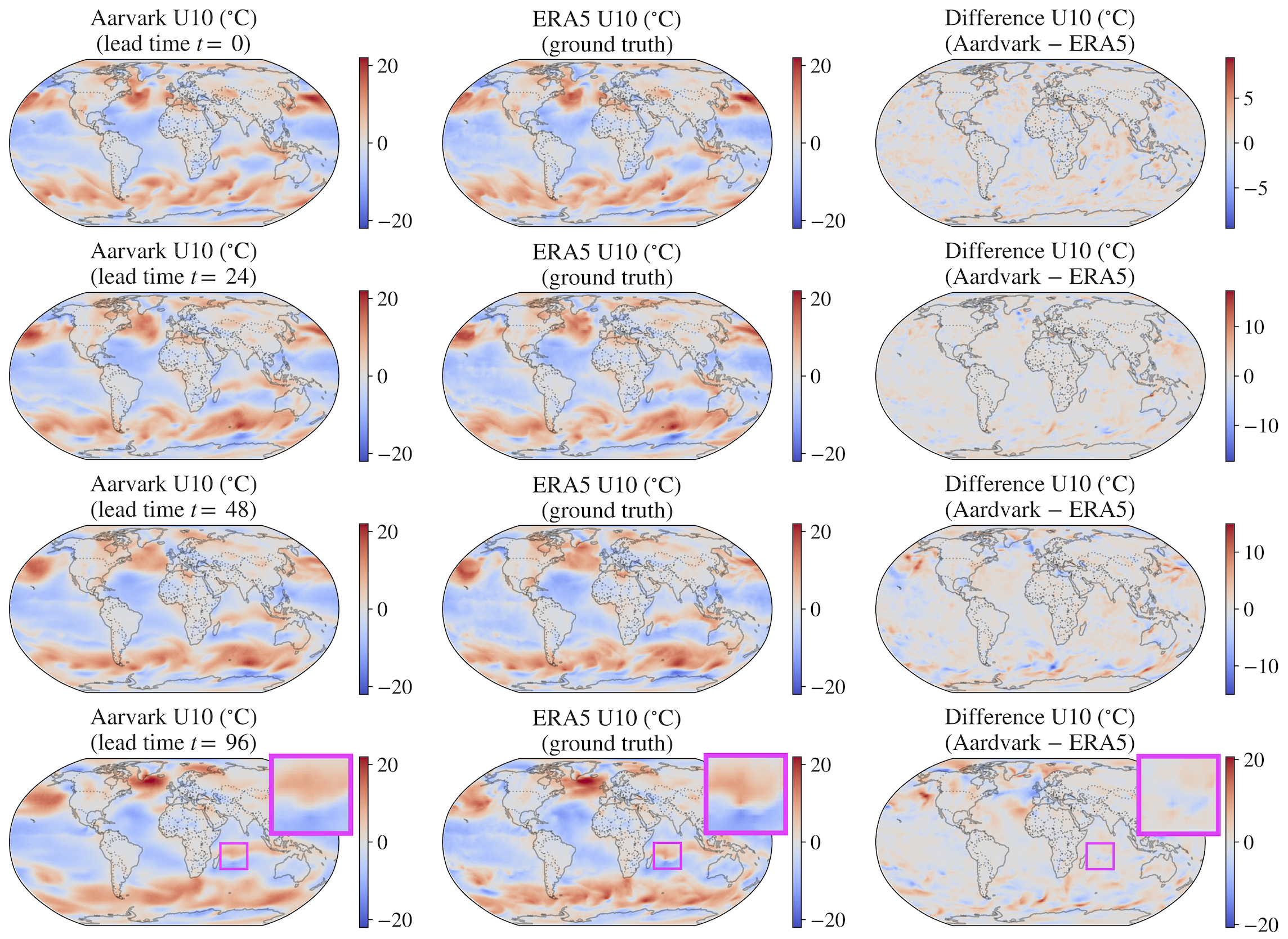}
\caption{
    \textbf{Illustration of Aardvark's global gridded forecasts for 10-metre wind speed.}
    Plots of the initial condition (first row) and subsequent forecasts (second, third and fourth rows) for 10-metre wind speed (U10), showing Aardvark's prediction (left), the ERA5 ground truth (middle), and the difference between the two (right).
    Lead time $t = 0$ corresponds 00:00 on the $11^{\text{th}}$ of January 2018.
    Aardvark correctly predicts large-scale features for this variable, and correctly predicts the formation and positioning of the tropical cyclone Berguitta (highlighted in the magenta boxes), which reached peak intensity on the $15^{th}$ of January 2018 off the coast of Madagascar.
    We emphasise that the model makes these predictions entirely from raw observations, without any NWP products as input.
    Appendix \ref{app:plots} gives example plots for all variables predicted by Aardvark}
\label{fig:initial_condition}
\vspace{-4mm}
\end{figure}

\begin{figure}[t!]
\centering
\includegraphics[width=\linewidth]{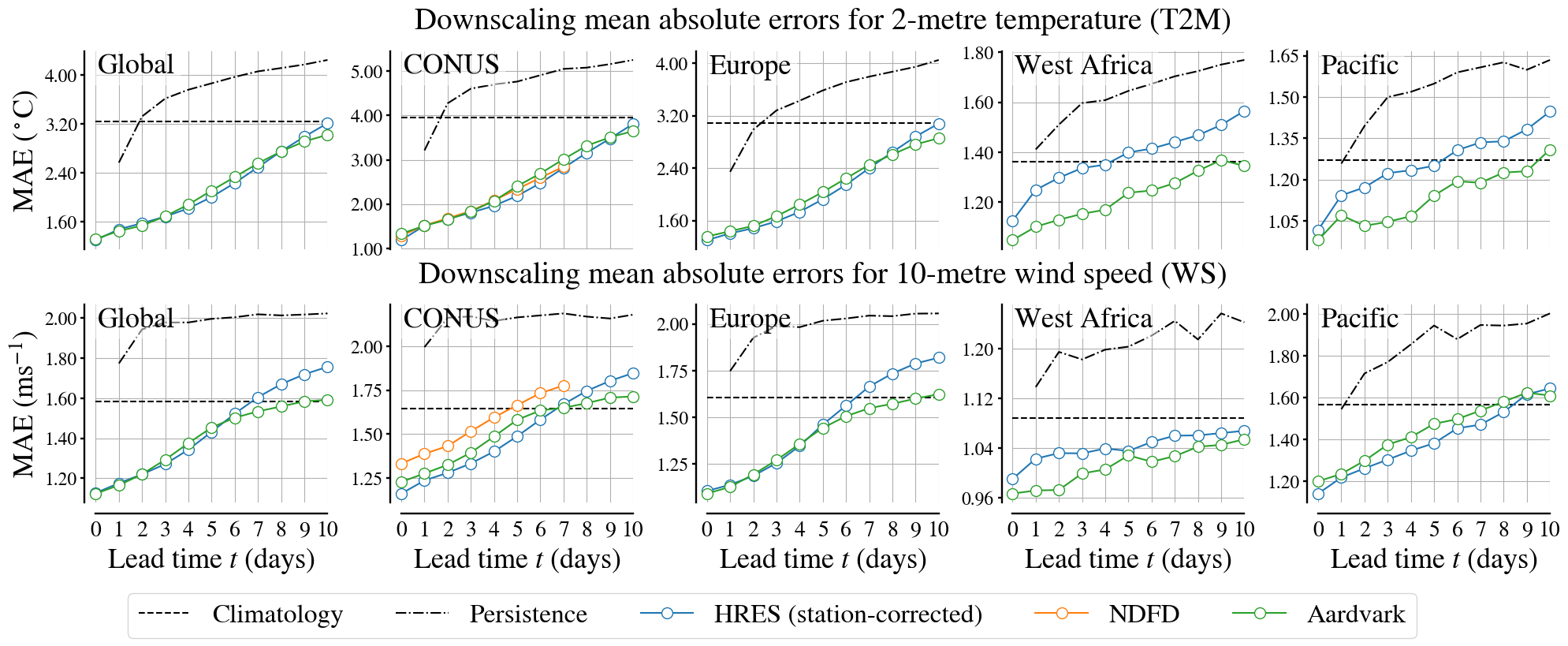}
\includegraphics[width=\linewidth]{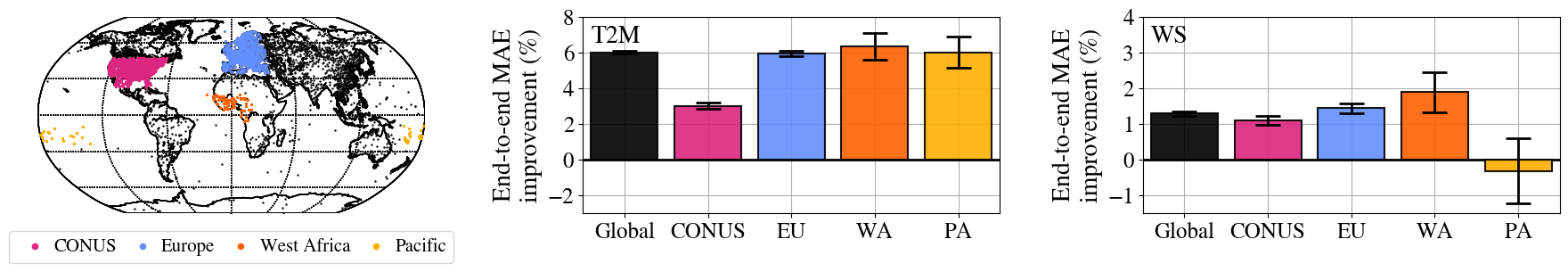}
\caption{
    \textbf{Station downscaling and end-to-end performance.}
    Results for station forecasting (top) and end-to-end optimisation (bottom) for the held-out test set (2018) of HadISD data.
    Here, Aardvark makes predictions at spatial locations observed during training, on temporally held out data, however can generate predictions at any arbitrary station location.
    For station forecasting (top) we compare Aardvark's  forecasts to two state-of-the-art NWP baselines, the National Digital Forecast Database (NDFD) for CONUS, and a version of HRES that we correct using a scale and bias term learned separately for each station (see text for discussion).
    In end-to-end fine-tuning (bottom), we compare the predictions of Aardvark for lead time $t = 1$ to those of its end-to-end fine-tuned counterpart for 2-metre temperature (T2M) and 10-metre wind speed (WS).
    We report the mean \% improvement in each variable by region (see bottom left) with 95\% confidence intervals.
    "Global" includes all stations (black and coloured)}
\label{fig:downscaling_and_end_to_end}
\end{figure}

\subsection*{Performance evaluation: station forecasting}
\label{subsec:results:downstream}

In the next stage of the weather prediction pipeline, global gridded forecasts are used as input to downstream models to produce a variety of products for end users.
One such category of products is producing local forecasts.
We focus on applying Aardvark Weather to predict 2-metre atmospheric temperature and 10-metre wind speed at off-the-grid station locations.
Accurate local predictions of temperature are vital for protection of public health during heatwaves \citep{kovats2006heatwaves,mayrhuber2018vulnerability} and cold waves \citep{huynen2001impact}, in addition to agriculture and other use cases.
Similarly, wind speed forecasts have a variety of end users for example in wind energy \citep{wang2016quantifying}, marine forecasting and fire weather forecasting.
We note that modules for any desired downstream product could be substituted for this station forecasting module. \\

There are significant differences in how agencies in different countries produce forecasts for end users.
In well resourced countries, for example the United States, station forecasts are produced using global models followed by higher resolution regional models out to a few days lead time and statistical post-processing \citep{usa2013}.
In contrast, in less well resourced areas while agencies have access to global products such as HRES and GFS they often do not have the computational resources and expertise to run local NWP at higher resolution or post-process the outputs \citep{powerbi2024}.
With these considerations in mind, in addition to reporting Aardvark's performance across all stations globally, we also report performance in four regions of particular interest: the contiguous United States (CONUS), Europe, West Africa and the Pacific.
These regions are illustrated in \Cref{fig:downscaling_and_end_to_end} (bottom left).
The United States and most European countries have well-resourced centres running local NWP for shorter lead times in addition to sophisticated post-processing of both global and local products. In contrast, West Africa and the Pacific are regions in which many centres are less well equipped.
Although some agencies in countries in these regions run sophisticated NWP pipelines, others utilise solely raw HRES forecasts and issue operational forecasts only for very short lead times \citep{powerbi2024}. \\

As the full end-to-end pipeline differs significantly for different meteorological agencies, a range of baselines could be used for station forecasts.
In addition to simple persistence and climatology, we opt to compare against two challenging baselines.
The first of these is a downscaled version of HRES.
This is created by first generating station forecasts from $0.25^{\circ}$ HRES by selecting the nearest gridpoint.
As post-processing adds significant value to global models \citep{bouallegue2023statistical} we subsequently learn an affine correction (a scale and a constant bias) on a per-station basis to correct for systematic biases.
This bias correction significantly improves the performance of HRES, especially over West Africa and the Pacific, and creates a strong baseline which we refer to as station-corrected HRES.
Second, over CONUS, we also compare against a full operational end-to-end baseline, the National Digital Forecast Database (NDFD) from the National Weather Service.
NDFD forecasts are an archive of data from National Weather Service offices produced by combining the output of multiple global and regional forecasting models, together with post-processing and input from human forecasters \citep{glahn2003new}. \\

\Cref{fig:downscaling_and_end_to_end} shows the performance of Aardvark, reported by variable and region.
Globally Aardvark generates skillful forecasts for both temperature and wind speed up to a lead time of 10 days, performing competitively with station-corrected HRES.
For temperature, Aardvark is competitive with station-corrected HRES over both CONUS and Europe.
In addition, remarkably, Aardvark matches the performance of the full operational NDFD baseline over CONUS.
For lower resource areas in West Africa and the Pacific Aardvark outperforms station-corrected HRES at all lead times.
For 10-metre wind speed, Aardvark has higher errors than station-corrected HRES over CONUS, and significantly outperforms the NDFD baseline.
Over Europe, Aardvark has similar errors with station-corrected HRES up to four days lead time, and outperforms it thereafter.
Finally, Aardvark generally outperforms station-corrected HRES over West Africa, while performing slightly worse over the Pacific.

\subsection*{End-to-end tuning}
\label{subsec:results:downstream}
End users of NWP products typically have a particular region and set of applications that are of interest. 
A powerful capability of Aardvark is the ability to tune the entire pipeline end-to-end to directly optimise for any desired quantity and region of interest. Optimising the performance for a particular end-user product would be challenging and expensive in a conventional NWP system.
To explore this capability, we fine-tune Aardvark to optimise predictions of 2-metre temperature and 10-metre wind speed at one day lead time globally and for each of the four regions. \\

We observe that fine-tuning Aardvark yields improvements both globally, as well in the specific regions of CONUS, Europe, West Africa and the Pacific (\cref{fig:downscaling_and_end_to_end}; bottom).
For temperature, fine-tuning Aardvark results in large reductions in mean absolute error (MAE) of 6\% over Europe, West Africa, the Pacific, and globally, and an improvement of 3\% over CONUS.
For 10-metre wind speed, small but statistically significant improvements of 1-2\% are observed for all regions except the Pacific.
To put these improvements into context, the last cycle update of the IFS led to improved scores on surface variables in the range of two to six percent and took over a year of development by a large team of scientists.

\section*{Discussion}
\label{sec:discussion}

We have introduced Aardvark Weather, an end-to-end weather forecasting system which is the first data-driven system to tackle the entire NWP pipeline.
Aardvark provides accurate forecasts that are orders of magnitude cheaper than any existing system including those that leverage recent AI forecasting models in place of numerical solvers, outperforming operational NWP systems on multiple tasks.
Generating a full forecast from observational data takes approximately one second on four NVIDIA A100 GPUs, compared to the approximately 1,000 node-hours required by HRES to perform data assimilation and forecasting \citep{buizza2018development} alone, before accounting for downstream local models and processing.
Learning an end-to-end model offers the additional capability to optimise the system to maximise performance over an arbitrary variable or region of interest, opening the door for the creation of cheap, bespoke models for any region globally.  \\

End-to-end forecasting will have significant potential for real world impact. Compared to conventional NWP systems, machine learning systems are not only faster and computationally cheaper, but are also significantly easier to improve and maintain. 
In conventional NWP a new module, for example for a novel parameterisation or micro-physics scheme, may take a team considerable time to build and integrate into the model.
End-to-end data-driven systems such as Aardvark elegantly bypass this issue using a single model in place of this complex pipeline.
The simplicity of this system both makes it both easier to deploy and maintain for users already running NWP, and also opens the potential for wider access to running bespoke NWP in areas of the developing world where agencies often lack the resources and expertise to run conventional systems.
There is also significant potential in the demonstrated ability to fine-tune bespoke models to maximise predictive skill for specific regions and variables. 
This capability is of interest to many end users in areas as diverse as agriculture, renewable energy, insurance and finance.\\

The results presented in this study only scratch the surface of the potential of Aardvark weather and end-to-end data-driven weather forecasting systems more broadly. 
Improvements to this system could be made through adding more input data modalities, increasing the observation window for the encoder module and utilisation of higher resolution and more sophisticated architectures for all three modules. This paradigm is easily extended to provide multiple other forecast modalities, including adding more variables and levels to the global forecasts
It also supports the addition of a diverse range of decoder modules to provide different types of end user forecasts such as hurricane, flood, severe convection, fire weather and other extreme weather warnings, and is equally applicable to longer lead times generating seasonal forecast products. 
Furthermore, additional observational modalities would allow for modelling of other components of the earth system, such as atmospheric chemistry for air quality forecasts and ocean parameters for marine forecasts.
We envision that Aardvark Weather will be the first of a new generation of end-to-end weather forecasting systems tackling these diverse tasks.\\

\section*{Methodology}
\label{sec:methodology}
\subsection*{Datasets: state estimation inputs}
We select multiple remote sensing and in-situ observations for input to the atmospheric state estimation module. To ensure that no NWP system is required for operational deployment of Aardvark, we select only data that are available at either level 1B or 1C processing level. Other requirements for inclusion of datasets are that they are available from 2007-2020 and are available in near real time to facilitate anticipated operational deployment. Where available for remote sensing products we utilise fundamental climate data records, where data from earlier generation sensors are homogenised to match the characteristics of current sensors, creating a consistent data record for training.\\

In-situ observations are included from land stations, marine platforms and radiosondes. 
In-situ land station observations measuring surface temperature (8719 stations), pressure (8016 stations), wind (8721 stations) and dew point temperature (8617 stations) at six hourly intervals are taken from the HadISD dataset \citep{dunn2012hadisd,dunn2016expanding}.
Marine in-situ observations are taken from the International Comprehensive Ocean-Atmosphere Data Set (ICOADS) \citep{freeman2017icoads} dataset.
This dataset consists of observations from ships and buoys globally, from which five variables are included, namely 2-metre air temperature, 10-metre northward and eastward wind, sea surface temperature and mean sea level pressure. As observations are not taken precisely on the hour, all observations from $t = -1$ hours to $t = 0$ are included in the input. Upper atmosphere observations of humidity, wind, geopotential and temperature are taken from The Integrated Global Radiosonde Archive (IGRA) dataset \citep{durre2006overview}.
This dataset consists of radiosonde observations at 1375 sites globally.
Each record contains observations at multiple levels, of which we select observations at the surface and 200, 500, 700 and 850hPa pressure levels. All profiles retrieved within the past six hours, from $t = -6$ hours to $t = 0,$ are included in the input.\\

As in-situ observations are limited in geographic coverage, remote sensing observations are included from scatterometers and microwave and infrared sounders. Input data from satellites are ingested in the form of level one granules each containing a six minute slice of observations or orbits. Although in principle the Aardvark Weather system can handle this data in its raw form, for simplicity in preliminary experiments data were first transferred to a regular one degree grid by nearest neighbour interpolation where the most recent observation is maintained in cases where multiple observations are available for the same gridpoint. \\

Several scatterometers are currently operational worldwide, of which we use the Advanced Scatterometer (ASCAT) instrument \citep{gelsthorpe2000ascat} aboard MetOp-A, B and C. ASCAT provides a triplet of three measurements of backscatter ($\sigma_0$) from which operational centres retrieve the wind speed and direction, using a geophysical model function which solves for the two unknowns as a function of the $\sigma_0$ triplet together with satellite metadata \citep{stoffelen2017cmod7}.
In contrast to this approach we opt to simply include the raw $\sigma_0$ values together with the metadata as channels to the encoder module, eliminating the complexity of the retrieval process. As all MetOp satellites are in low earth orbit (LEO), with a revisit time of approximately 24 hours, the input to the state estimation  module comprises of the latest ASCAT observations available within the grid box from any of the three platforms on a regular $1^{\circ}$ longitude-latitude grid from $t = -1$ days to $t = 0$ days. \\

In operational NWP, temperature and humidity profiles in the upper atmosphere are retrieved using infrared and microwave sounder instruments \citep{rosenkranz2001retrieval}. For this purpose we include the Advanced Microwave Sounding Units A and B (AMSUA \& AMSUB), and the Microwave Humidity Sounder (MHS)  instruments for microwave observations and the High Resolution Infrared Radiation Sounder (HIRS/4) for infrared observations. Together, these instruments comprise the Advanced TIROS Operational Vertical Sounder (ATOVS) system used operationally to retrieve temperature and moisture profiles \citep{li2000global}. Observations for AMSU-A, AMSU-B, MHS and HIRS are taken from the NOAA 15 through 19, Aqua and MetOp-A satellites. In operational NWP systems, both retrieved profiles and raw radiances are assimilated. Similar to ASCAT, profiles of the target variable are retrieved using a geophysical model function taking in the raw radiances and satellite metadata and solving for the desired observational profiles. We again opt to input the raw radiances together with the satellite metadata directly into the state estimation module without relying on higher level retrievals.
As for ASCAT, the dataset consists of the latest observations from $t = -1$ to $t = 0$ days, taken within a grid box of a regular $1^{\circ}$ longitude-latitude grid. \\

We augment the ATOVS observations with data from a hyperspectral infrared sounder, the Infrared Atmospheric Sounding Interferometer (IASI). IASI captures data at a much higher spectral resolution than HIRS/4, with a total of 8461 channels across three bands. To limit input data volume, we take the leading 15 principal components across these channels, a technique demonstrated to lead to limited performance degradation in operational NWP systems. \\

While platforms carrying scatterometer and passive microwave sounder instruments in LEO provide high-resolution observations, they have the disadvantage of lower temporal resolution.
In contrast, geostationary satellites provide very high temporal resolution though with more limited instrumentation.
As the available channels on geostationary satellites vary geographically and with time, we opt to use a composite product, the Gridded Satellite dataset \citep[GridSat;][]{knapp2018gridded}, which provides homogenised retrievals of infrared and vapour window channels over standard geostationary platforms. For this data source we include the image taken at $t=0$\\

To account for diurnal, seasonal and longer term variations in the data, we include temporal information as input both to the encoder and forecasting modules. These channels consist of $\sin(\frac{2\pi d}{366})$, $\cos(\frac{2\pi d}{366})$, $\sin(\frac{2\pi h}{24})$ and $\cos(\frac{2\pi h}{24})$ where $d$ is the day of year and $h$ the hour of day. The absolute year is also included to account for any changes in data characteristics over the training record. In order to account for the effects of orography on the weather system, we include several sources of orographic information taken from the ERA5 dataset \citep{hersbach2020era5} as static fields. These are the geopotential at surface level, angle of sub-grid scale orography, anisotropy of sub-grid scale orography, slope of sub-grid scale orography and standard deviation of orography.

\subsection*{Datasets: pre-training}
The modular structure of Aardvark leverages ERA5 reanalysis data  during the training phase to increase the length of the data record available.
ERA5, or the Fifth Generation of the European Centre for Medium-Range Weather Forecasts (ECMWF) Reanalysis \citep{hersbach2020era5}, is a state-of-the-art global atmospheric reanalysis dataset.
It provides comprehensive information on various meteorological parameters such as temperature, humidity, wind, and geopotential, covering the period from 1940 to present.
From this we elect to train on data from 1979 onwards, coinciding with the beginning of widely available remote sensing observations which significantly improves the quality of the atmospheric reanalysis product.

\subsection*{Datasets: baselines}

For the global gridded forecast experiments we compare the performance of Aardvark against four baselines: persistence, climatology, HRES and GFS.
Persistence and climatology provide simple baselines for assessing whether a forecasting system is skillful. In persistence forecasting, it is assumed that the weather will remain unchanged from $t = 0$ at all future lead times.
For the climatology baseline we utilise the climatology product from WeatherBench 2 \citep{rasp2023weatherbench}. Here, the predicted state is obtained by taking the mean value of all ERA5 observations from 1990-2017 for a given day of the year and hour using a sliding window of length 61 days. \\

The IFS and GFS are the two most widely used global operational NWP systems. 
As the focus of this study is on deterministic study, we choose to compare our results to the HRES and GFS, deterministic runs at a resolution of $0.10^{\circ}$ degrees and $0.25^{\circ}$ degrees respectively. 
These constitute extremely challenging baselines for comparison to Aardvark Weather which operates at a $1.41^{\circ}$ resolution with just five vertical levels. For comparison to Aardvark, HRES and GFS outputs are interpolated to $1.41^{\circ}$ resolution. \\

For station forecasts we consider four baselines. Persistence and climatology are calculated based on station observations. For 2-metre temperature we calculate daily climatology, and for 10-metre wind speed monthly. We further consider two more challenging baselines: station-corrected HRES and NDFD over the contiguous United States. As HRES is a gridded product, sub-grid scale processes are not resolved. We therefore learn a bias correction individually for each station on the 2007-2017 training set, and use this to correct data in the 2018 test set. NDFD is produced by the National Weather Service in the United States of America, and is a state-of-the art local forecasting system \citep{NWS2024}. Forecasts in the NDFD are created from an ensemble of over 30 models \citep{NWSModels2024}, including the IFS and GFS together with high resolution regional models at shorter lead times. The data from these systems is the shown to human forecasters at different NWS offices who create the final forecast. Our station forecasts are taken as the nearest gridbox forecast from the final NDFD forecast which is at approximately 2km resolution. NDFD therefore constitutes an extremely challenging baseline, capturing the full complexity of operational forecasting pipeline. 

\subsection*{Evaluation metrics}
For the global gridded forecasting experiments we compare models on latitude weighted root mean squared error.
Given arrays of gridded target forecasts $y$ and gridded target predictions $\hat{y},$ the latitude weighted RMSE of variable $v$ is calculated as
\begin{equation}
\text{LW-RMSE}(y, \hat{y}, v) = \frac{1}{B}\sum_{b=1}^B \sqrt{\frac{1}{HW}\sum_{h = 1}^H\sum_{w = 1}^W \alpha_h (y_{bhwv} - \hat{y}_{bhwv})^2} \label{eq:lwrmse}
\end{equation}
where $b$ indexes batch elements, $v$ indexes atmospheric variables, $h$ and $w$ index latitude and longitude coordinates and $\alpha_h$ are the latitude weights, defined as
\begin{equation}
\alpha_h = \frac{\cos{\theta_h}}{\frac{1}{H}\sum_{h=1}^H \cos{\theta_h}}, \label{eq:rmse}
\end{equation}
where $\theta_h$ is the latitude along the latitude-wise index $h,$ so that their average is equal to one.
For the station forecasting experiments we compare methods on root mean squared error.
Given arrays of station target temperatures $y$ and predictions $\hat{y},$ the RMSE is calculated as
\begin{equation}
\text{RMSE}(y, \hat{y}) = \frac{1}{B}\sum_{b=1}^B \sqrt{\frac{1}{N}\sum_{n = 1}^N (y_{bn} - \hat{y}_{bn})^2}
\end{equation}
where $b$ indexes batch elements, and $n$ indexes the stations in the forecast.

\subsection*{Training objectives}
Separate training objectives are utilised for each of the three modules.
For all three modules, we normalise the targets on a per-variable basis.
We note that the encoder and processor modules, which involve multiple target variables, this normalisation has the effect of implicitly weighting the variables, due to the scaling applied during normalisation.
For the encoder module we determine an additional weighting by first training the model with using a latitude weighted RMSE objective of the form
\begin{equation} \label{eq:sumlwrmse}
\text{SUM-LW-RMSE}(y, \hat{y}) = \frac{1}{V} \sum_{v=1}^V \text{LW-RMSE}(y, \hat{y}, v).
\end{equation}
In this initial run, all variables are therefore weighted equally.
Next, weights $\beta_v$ are produced for each variable by taking the reciprocal of the latitude weighted RMSE for each variable multiplied by a factor of three to generate weights within the range of approximately 0 to 1.
The training objective for the encoder uses these weights, giving the variable and latitude weighted RMSE
\begin{equation}
\text{VLW-RMSE}(y, \hat{y}) = \frac{1}{V} \sum_{v = 1}^V \beta_v \, \text{LW-RMSE}(y, \hat{y}, v).
%
\end{equation}
For the processor module the training objective is the SUM-LW-RMSE (eq. \ref{eq:sumlwrmse}).
However, the processor module is trained to predict residuals (see ``Processor module'' below).
We found the implicit weighting that is applied via normalisation works well, and we did not further weight the variables individually.
Finally, for the decoder module the training objective is the same as for evaluation, that is \cref{eq:rmse}.

\subsection*{Model architecture}

Aardvark Weather is a neural process model \citep{garnelo2018conditional}. Neural processes are a family of deep learning models that provide a flexible framework capable of learning with off-the-grid data, missing and sparse data and providing probabilistic predictions at arbitrary locations at test time. These characteristics are ideally suited to working with complex environmental data for example in climate downscaling and sensor placement \cite{andersson2023environmental,markou2022practical,vaughan2022convolutional,vaughan2021multivariate,bruinsma2023autoregressive}. \\

Our specific architecture is a novel member of the neural process family combining SetConv layers developed for the convolutional conditional neural process \citep{gordon2019convolutional}, which handle off-the-grid and sparse data modalities and produce off-the-grid predictions, together with a vision transformer backbone currently used in state-of-the-art AI-NWP forecasting systems \citep{bodnar2024aurora}. This provides scalability not currently attainable with standard transformer neural process models with attention based encoders \citep{nguyen2023scaling}, whilst still retaining the flexibility to handle diverse data modalities. 
Here we give details for the architectures of these modules, how they are trained and fine-tuned, and how they are deployed.
In the discussion that follows, note that the encoder, processor and decoder modules all receive auxiliary channels, such as temporal embeddings and orographic information, as input.
For simplicity, we suppress these channels in our exposition, but it should be understood that all three modules receive them as input.

\subsection*{Encoder module}
The encoder module $E$ takes raw observations as input and outputs a gridded estimate of the initial state of each variable for the processor module.
Let $o_\tau = \{o_{\tau, 1}, \dots, o_{\tau, N}\}$ be the set of observations corresponding to time $\tau,$ where each $o_{\tau, N}$ corresponds to the observations of a single data modality.
Each $o_{\tau, n} = (x_{\tau, n}, y_{\tau, n})$ consists of a set of observations $y_{\tau, n}$ and their corresponding longitude and latitude coordinates $x_{\tau, n}$.
Each data modality is either on-the-grid or off-the-grid, and has a corresponding function $\psi_n$ to transform $o_{\tau, n}$ into a gridded representation of fixed dimensionality.
For gridded observations, $\psi_n$ consists of the addition of a masking channel to distinguish missing data from observed data in the grid, followed by bilinear interpolation to a regular $1.41^\circ$ grid.
For off-the-grid observations, each $\psi_n$ consists of a SetConv layer \citep{gordon2019convolutional} with a learnable length scale.
The SetConv layer produces a gridded representation of the data, as well as an accompanying density channel which carries information about the presence or absence of data, to handle irregularly sampled observations.
The regular gridded representations of the modalities are concatenated to give a single gridded representation of dimension $C \times H \times W,$ where $C$ is the number of resulting channels, $H$ is the number of latitude points and $W$ is the number of longitude points.
This representation of the input data is fed into the backbone of the module, consisting of a vision transformer $V_e$ with patch size three, eight transformer blocks and latent dimension 512.
Embeddings for each patch use an MLP following \citep{dosovitskiy2020image}.
The encoder outputs the initial state estimate $\hat{s}_{\tau,0}$ at time $\tau$ with dimension $ 24 \times W \times H$ where 24 is the number of variables modelled in the forecasting module.
Putting this together, we have
\begin{equation} \label{eq:encoder}
    \hat{s}_{\tau, 0} = E(o_{\tau}) = V_e\left(\odot_{n=1}^N \psi_n(o_{\tau, n})\right),
\end{equation}
where $s_{\tau, 0}$ is the estimated initial state corresponding to time $\tau$ and $\odot$ denotes concatenation.
The encoder module is trained to predict ERA5 reanalysis targets using the VLW-RMSE (eq. \ref{eq:vlwrmse}) as its loss function.
We train the module for 150 epochs using AdamW with early stopping and a cosine learning rate scheduler starting at an initial learning rate of $5 \times 10^{-4}$ and decaying to zero at the final epoch.

\subsection*{Processor Module}
The processor module $P$ takes the initial state estimate $s_{\tau, 0}$ as input and outputs forecasts for lead-times of one to ten days.
This module consists of ten separate vision transformers, $V_p^{(1)}, \dots, V_p^{(10)},$ which are composed to produce gridded global forecasts at each of the ten lead times we consider.
All vision transformers have a patch size 5, latent dimension 512 and 16 transformer blocks.
To improve modelling of interactions between variables we add cross-attention between variables at the start of the network, as suggested by \citep{nguyen2023climax}.
The processor is trained using a pre-training phase followed by a fine-tuning phase.
Let $s_{\tau, t}$ be the ERA5 state corresponding to time $t$ and lead time $\tau.$
During pre-training, the first vision transformer, \smash{$V_p^{(1)},$} is trained to ingest $s_{\tau, 0}$ as input and predict the residual $s_{\tau, 1} - s_{\tau, 0}$ using the SUM-LW-RMSE loss (eq. \ref{eq:sumlwrmse}).
We pre-train, \smash{$V_p^{(1)}$} for 100 epochs using AdamW with a cosine learning rate scheduler starting at an initial learning rate of $5\times10^{-4}$ and decaying to zero at epoch 100.
During the fine-tuning phase, we train each vision transformer to work with the estimated state produced by the previous transformer as follows.
Recall that $\hat{s}_{\tau, 0}$ is the estimated state produced by the encoder module.
First, we fine-tune \smash{$V_p^{(1)}$} to predict $s_{\tau, 1} - \hat{s}_{\tau, 0}$ using the initial state $\hat{s}_{\tau, 0}$ as input.
Once \smash{$V_p^{(1)}$} has been fine-tuned, we compute \smash{$\hat{s}_{\tau, 1} = \hat{s}_{\tau, 0} + V_p^{(1)}(\hat{s}_{\tau, 0}),$} and initialise the network \smash{$V_p^{(2)}$} using the weights of \smash{$V_p^{(1)}.$}
We then fine-tune \smash{$V_p^{(2)}$} to predict $s_{\tau, 2} - \hat{s}_{\tau, 1}$ using $\hat{s}_{\tau, 1}$ the previously estimated initial state as input.
We proceed sequentially in this fashion, until all networks have been initialised and fine-tuned.
We note that this procedure can be regarded as an instance of the \textit{pushforward trick} \citep{brandstetter2023message}.
At deployment time, we compose the transformers to obtain a forecast for the desired lead time, that is
\begin{equation} \label{eq:processor}
    s_{\tau, t} = P(s_{\tau, 0}, t) = \tilde{V}_p^{(t)} \circ \dots \circ \tilde{V}_p^{(1)}(s_{\tau, 0}),
\end{equation}
where $\tilde{V}_p^{(t)}(\cdot) = \cdot + V_p^{(t)}(\cdot)$ and $s_{\tau, 0} = E\big(o_{\tau}\big)$ is the initial state produced by the encoder.


\subsection*{Decoder module}
The final step in the forecasting pipeline is the decoder module.
For each lead-time $t$, we train a lightweight convolutional station forecasting module $D_t,$ which takes the gridded estimated state $s_{\tau, t},$ as well as target longitude-latitude coordinates $x$ and auxiliary orographic information as inputs, and produces predictions for the corresponding station temperature measurements $y_{\tau, t}.$
Each $D_t$ consists of a UNet architecture \citep{ronneberger2015unet}, followed be a SetConv layer which maps on-grid predictions to predictions at arbitrary station locations, followed by an MLP which incorporates the auxiliary orographic information, to produce local forecasts $\hat{y}_{t, \tau}.$
The UNet consists of four encoder blocks (which consist of 2D convolutions, BatchNorm layers, ReLU activations and MaxPool operations) followed by four decoder blocks (which consist of transpose 2D convolutions, BatchNorm layers, ReLU activations and MaxPool operations).
The encoder and decoder blocks have skip connections, and channel dimensions $(16, 32, 64, 128, 64, 32, 16, 1)$. 
We train each $D_t$ for 10 epochs, using AdamW, with a learning rate of $1 \times 10^{-3}$ and the RMSE loss (eq. \ref{eq:rmse}).
To produce local forecasts at coordinates $x,$ we compute
\begin{equation} \label{eq:decoder}
    \hat{y}_{\tau, t} = D_t(s_{\tau, t}, x).
\end{equation}
where $s_{\tau, t}$ is the global forecast defined in \cref{eq:processor}.

\subsection*{End-to-end deployment}
At deployment time, no ERA5 input is required to run the system.
To obtain global forecasts, we compose the encoder and processor together, and compute
\begin{equation} \label{eq:e2e_deployment_global}
    \hat{s}_{\tau, t} = P_t \circ E(o_\tau)
\end{equation}
where $P_t(\cdot) = P(\cdot, t).$
If we want to produce local station forecasts, we compose the encoder, processor as well as decoder modules, and compute
\begin{equation} \label{eq:e2e_deployment_local}
    \hat{y}_{\tau, t} = D_t (P_t \circ E(o_\tau), x).
\end{equation}

\subsection*{End-to-end finetuning}
In order to perform end-to-end fine-tuning, we compose the encoder together with the lead time $t = 1$ processor and decoder modules, producing local station forecasts for lead time $t = 1$ given by
\begin{equation} \label{eq:e2e_finetuning}
    \hat{y}_{\tau, 1} = D_1 (P_1 \circ E(o_\tau), x).
\end{equation}
We then fine-tune all three networks jointly with either 2-metre temperature or 10-metre windspeed station observations $y_{\tau, 1}$ as the only targets, using the RMSE loss (eq. \ref{eq:rmse}).
We use AdamW and fine-tune the modules for 25,000 gradient steps with a constant learning rate of $5 \times 10^{-5}$ and early stopping. \\

In addition, for the end-to-end fine-tuned models, we perform region-based model selection during evaluation.
Specifically, we evaluate each of the model checkpoints generated during training on the validation data on the data from each of the regions we consider, namely Global, CONUS, Europe, West Africa and the Pacific.
For each region, we then select the best checkpoint and evaluate this on the test data corresponding to the given region.

\begin{Backmatter}

\paragraph{Acknowledgements}
This work was generously supported by The Alan Turing Institute, with funding and access to computational resources.
Anna Vaughan acknowledges the UKRI Centre for Doctoral Training in the Application of Artificial Intelligence to the study of Environmental Risks (AI4ER), led by the University of Cambridge and the British Antarctic Survey, and studentship funding from Google DeepMind.
Stratis Markou acknowledges funding from the Vice Chancellor's and George \& Marie Vergottis scholarship of the Cambridge Trust, and the Qualcomm Innovation Fellowship.
Will Tebbutt acknowledges funding from Huawei and EPSRC grant (EP/W002965/1).
James Requeima acknowledges funding from the Data Sciences Institute at the University of Toronto.
J.~Scott Hosking is supported by the Alan Turing Institute's Turing Research and Innovation Cluster in Digital Twins (TRIC-DT) and the Environment and Sustainability Grand Challenge, and EPSRC grant EP/Y028880/1. Richard E.~Turner is supported an EPSRC Prosperity Partnership grant EP/T005386/1 between the University of Cambridge and Microsoft.
We would like to thank Tomas Lazauskas for Cloud engineering support in setting up the compute platform, John Bronskill for technical advice on both compute and machine learning techniques, Peter Dueben for advice on baselines and Peter Lean for advice on counting the number of observations input to the IFS.

\paragraph{Competing Interests}
The authors do not have any competing interests to declare.

\paragraph{Data availability}
The dataset to run Aardvark Weather will be made available at https://huggingface.co/datasets/av555/aardvark-weather on the completion of peer review. All data required for peer review will be made available to reviewers.

\paragraph{Code availability}
Code to run Aardvark Weather will be made available at https://github.com/annavaughan/aardvark-weather-public on the completion of peer review. All code required for peer review will be made available to reviewers.

\paragraph{Author contributions}
A.V and R.T conceptualised the project. A.V, S.M, W.T, J.R, W.B and R.T designed the experiments. A.V selected and collected all data and designed the end-to-end system.  A.V, S.M and W.T implemented the codebase. A.V, S.M, W.T and R.T wrote the initial draft of the paper and S.M produced all figures. T.A, M.H, N.L, M.C, S.H and all aforementioned authors provided feedback on results at various stages of the project and contributed to the final version of the manuscript. 

\bibliographystyle{unsrt}
\bibliography{main}

\newpage
\section{Full global forecast results}
\label{app:forecast_full}
Here we provide full results for the latitude weighted RMSE performance of Aardvark together with our baselines, namely climatology, persistence, GFS and HRES, across all variables predicted by Aardvark.

\begin{figure*}[h!]
\centering
\includegraphics[width=\linewidth]{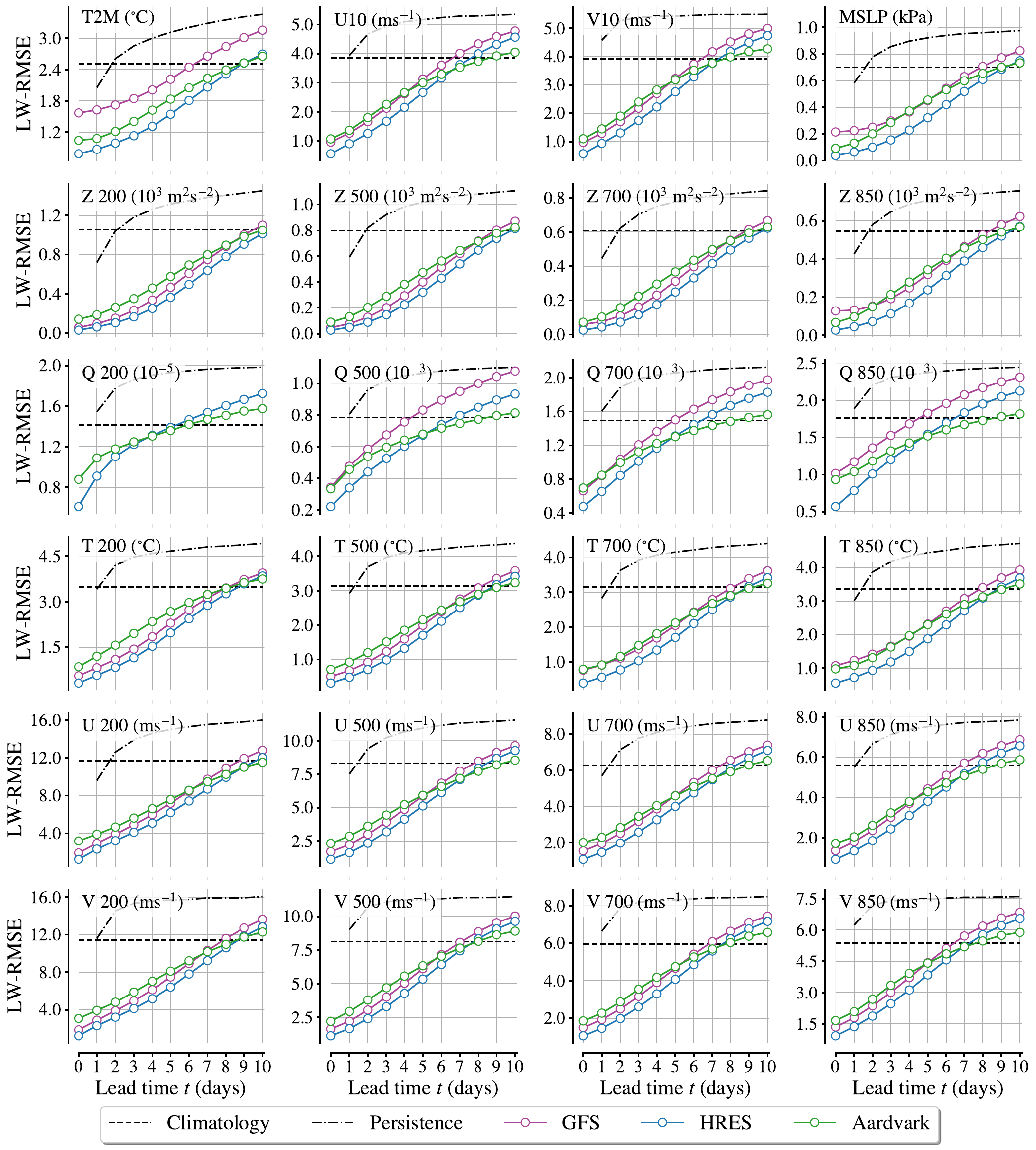}
\caption{
Latitude-weighted RMSE using ERA5 reanalysis data as the ground truth, on the held out test set (2018), for the four surface variables (first row) and all upper level variables (second row through to sixth row), as a function of lead time $t.$
At lead time $t = 0,$ Aardvark predicts the initial atmospheric state from \emph{from observational data alone}.
The error at $t = 0$ corresponds to the error in the initial state.
Note that HRES has non-zero error at $t = 0,$ as it is compared to ERA5 reanalysis ground truth}
\label{fig:forecast_full}
\end{figure*}

\newpage
\section{Additional initial state estimation and forecast plots}
\label{app:plots}
Here we include additional plots of estimated states and forecasts produced by Aardvark.
Lead time $t = 0$ corresponds to the estimated initial state.
We set $t = 0$ to be the $11^{\text{th}}$ of January 2018, corresponding to the formation of tropical cyclone Berguitta.

\begin{figure}[h]
    \centering
    \includegraphics[width=0.90\linewidth]{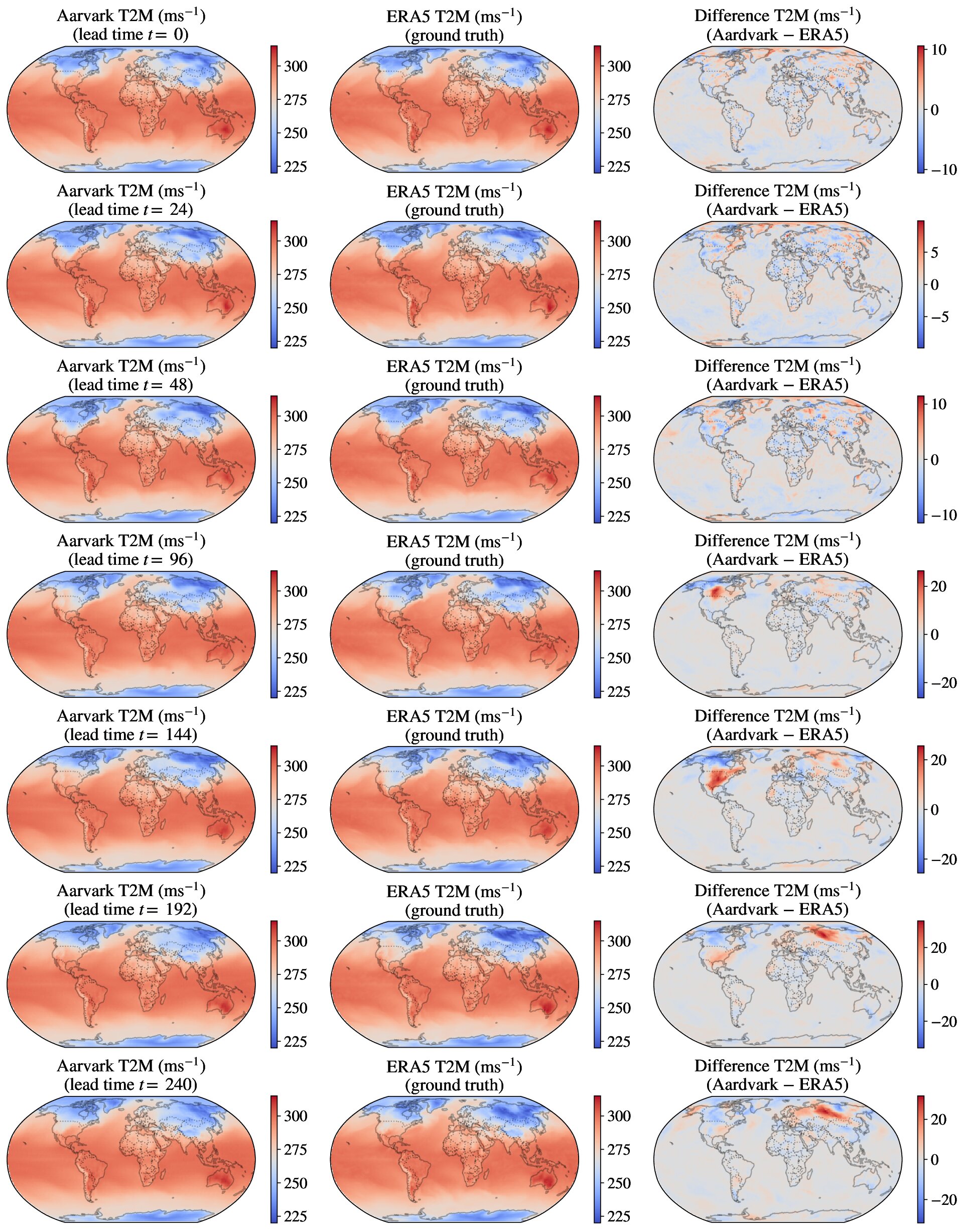}
    \caption{
        Illustration of the predictions of Aardvark.
        Note $t = 0,$ corresponds to $11^{\text{th}}$ of January 2018}
    \label{fig:app:0}
    \vspace{-5mm}
\end{figure}

\begin{figure}[h]
    \centering
    \includegraphics[width=0.90\linewidth]{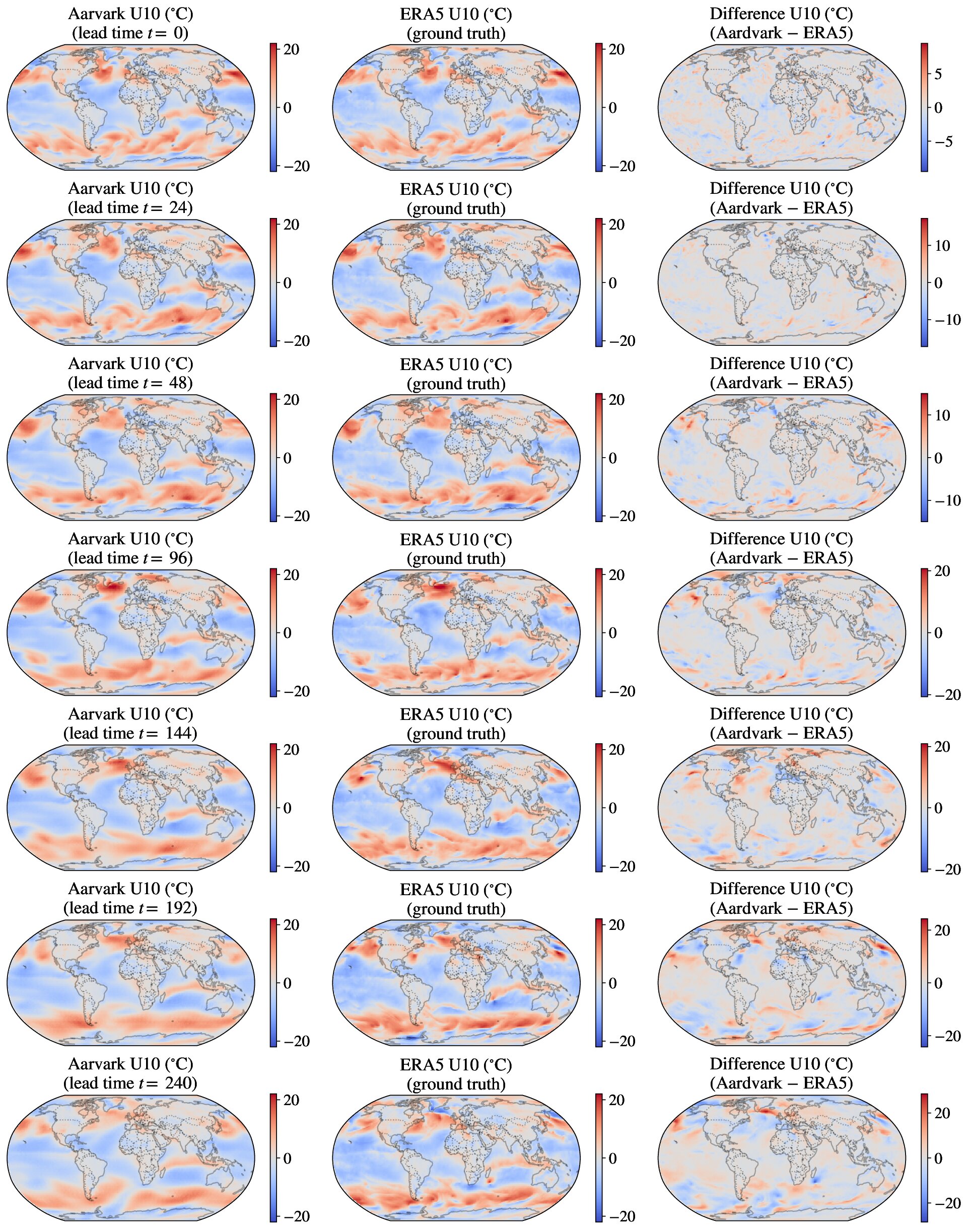}
    \caption{
        Illustration of the predictions of Aardvark.
        Note $t = 0,$ corresponds to $11^{\text{th}}$ of January 2018}
    \label{fig:app:1}
    \vspace{-5mm}
\end{figure}

\begin{figure}[h]
    \centering
    \includegraphics[width=0.90\linewidth]{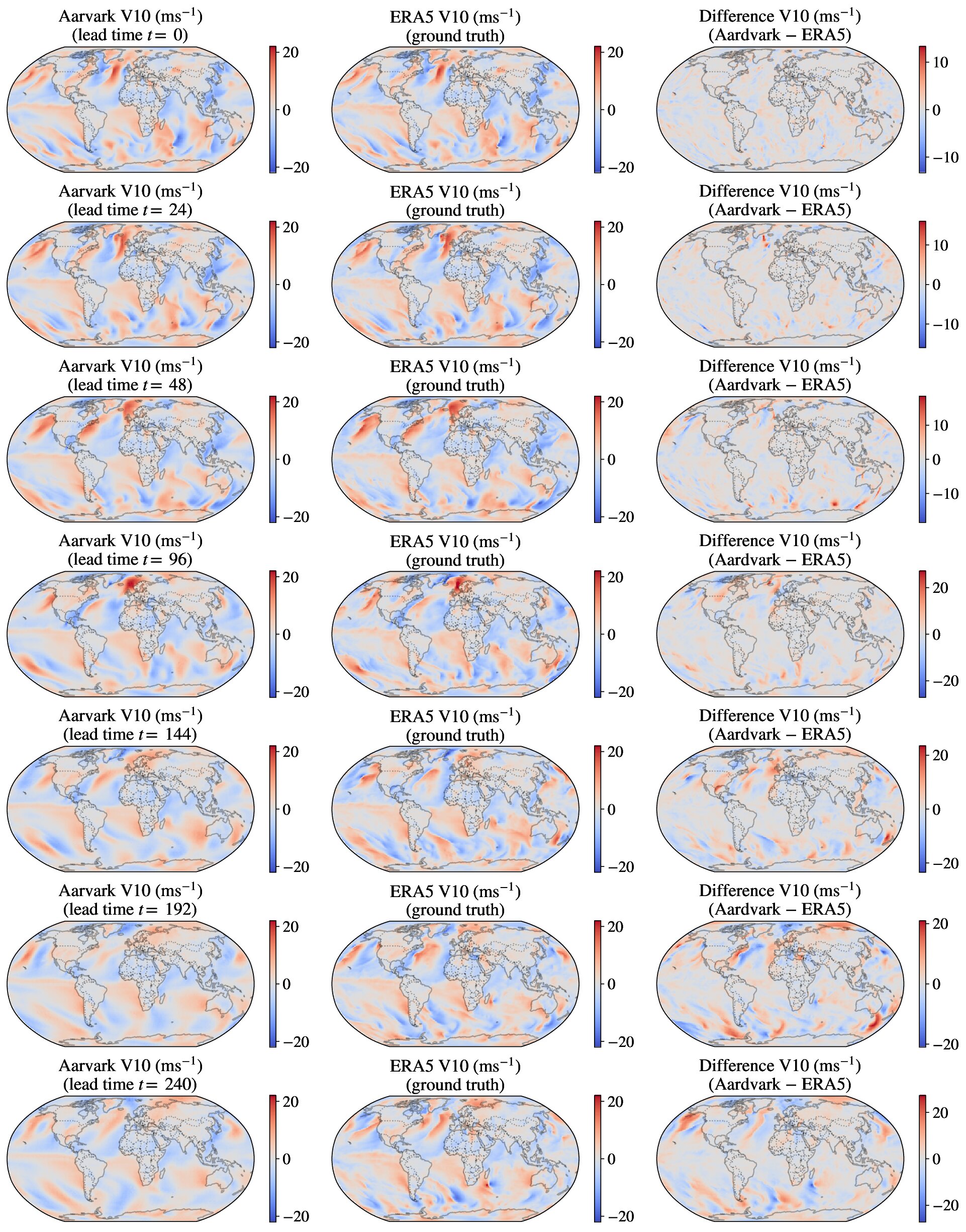}
    \caption{
        Illustration of the predictions of Aardvark.
        Note $t = 0,$ corresponds to $11^{\text{th}}$ of January 2018}
    \label{fig:app:2}
    \vspace{-5mm}
\end{figure}

\begin{figure}[h]
    \centering
    \includegraphics[width=0.90\linewidth]{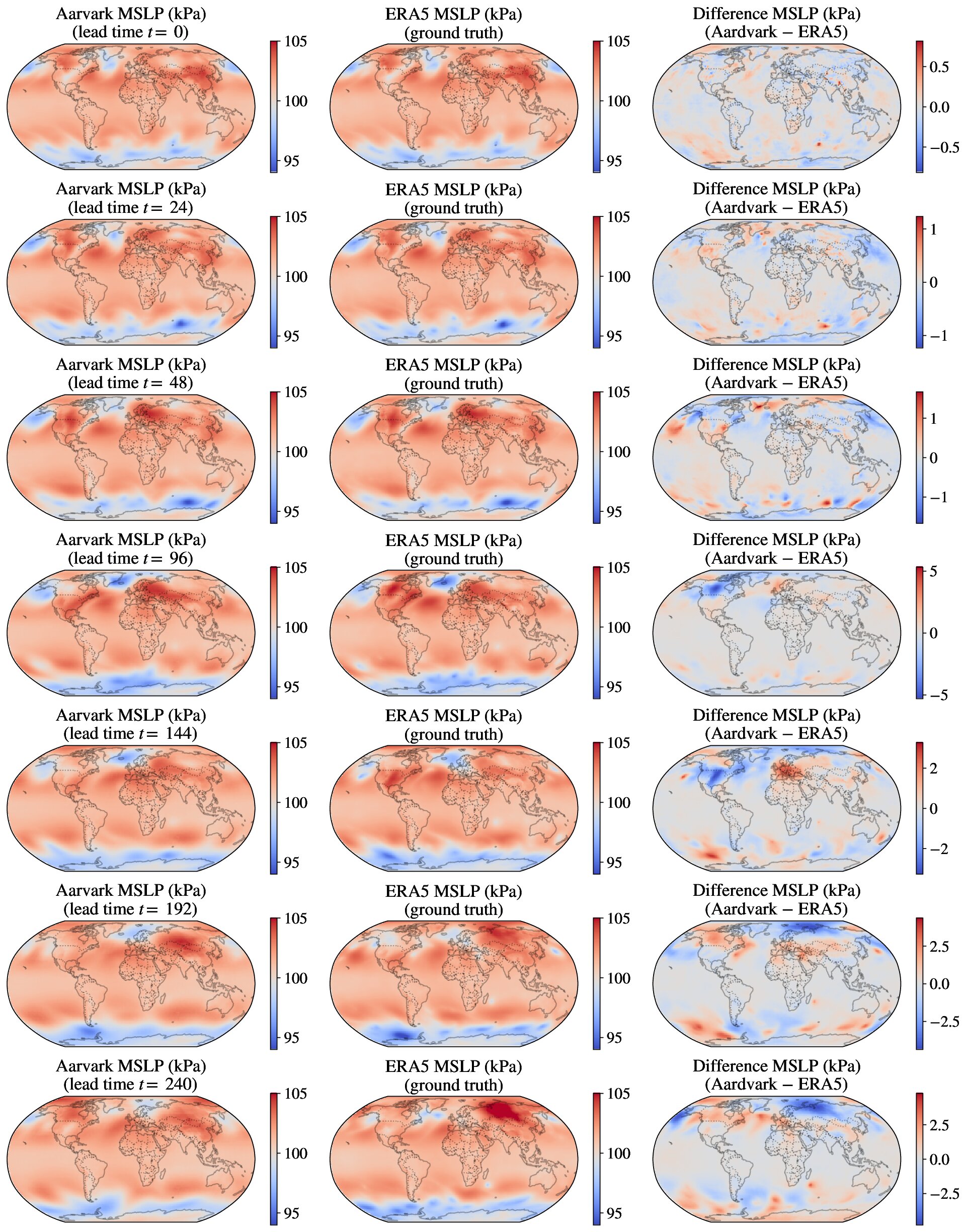}
    \caption{
        Illustration of the predictions of Aardvark.
        Note $t = 0,$ corresponds to $11^{\text{th}}$ of January 2018}
    \label{fig:app:3}
    \vspace{-5mm}
\end{figure}

\begin{figure}[h]
    \centering
    \includegraphics[width=0.90\linewidth]{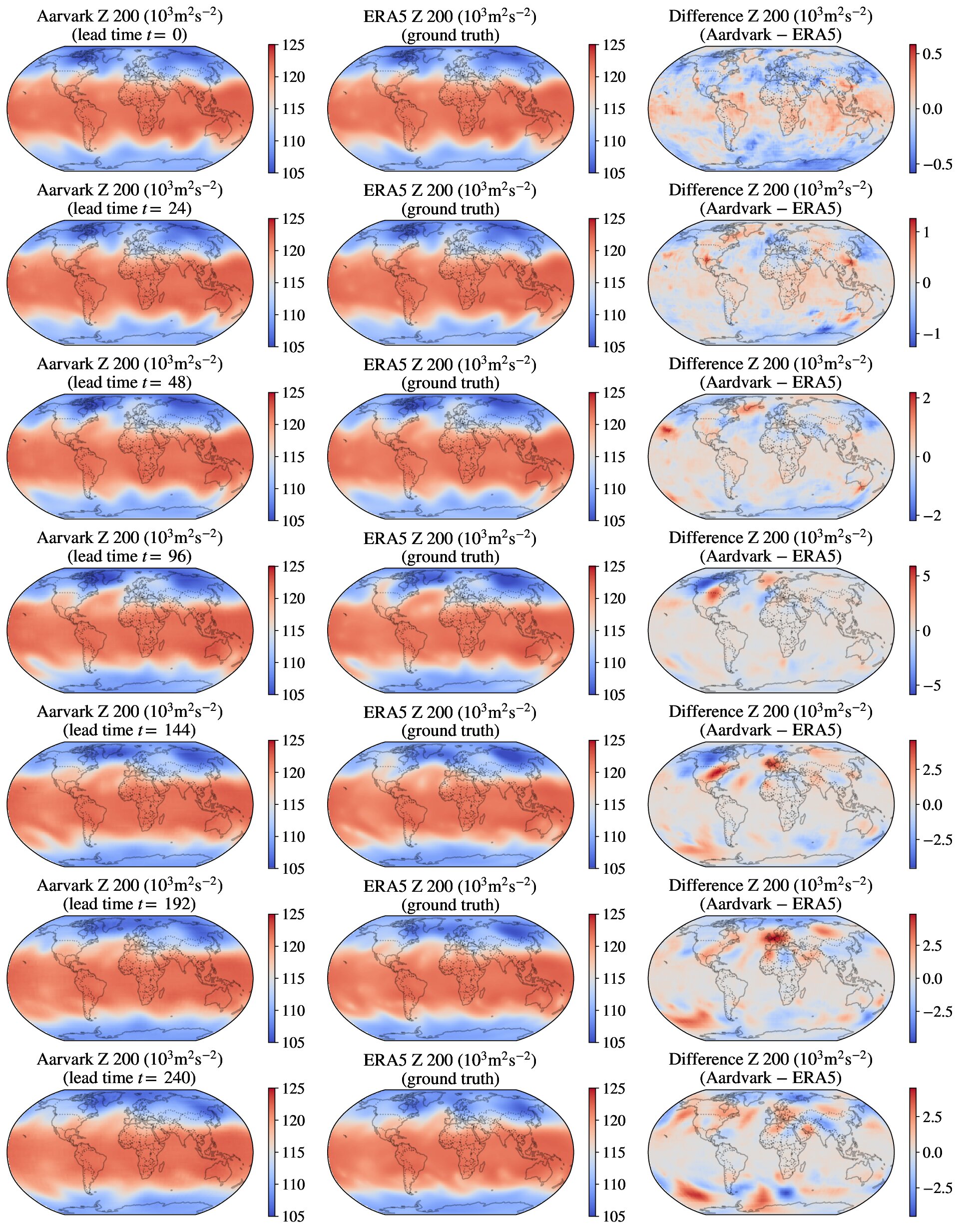}
    \caption{
        Illustration of the predictions of Aardvark.
        Note $t = 0,$ corresponds to $11^{\text{th}}$ of January 2018}
    \label{fig:app:4}
    \vspace{-5mm}
\end{figure}

\begin{figure}[h]
    \centering
    \includegraphics[width=0.90\linewidth]{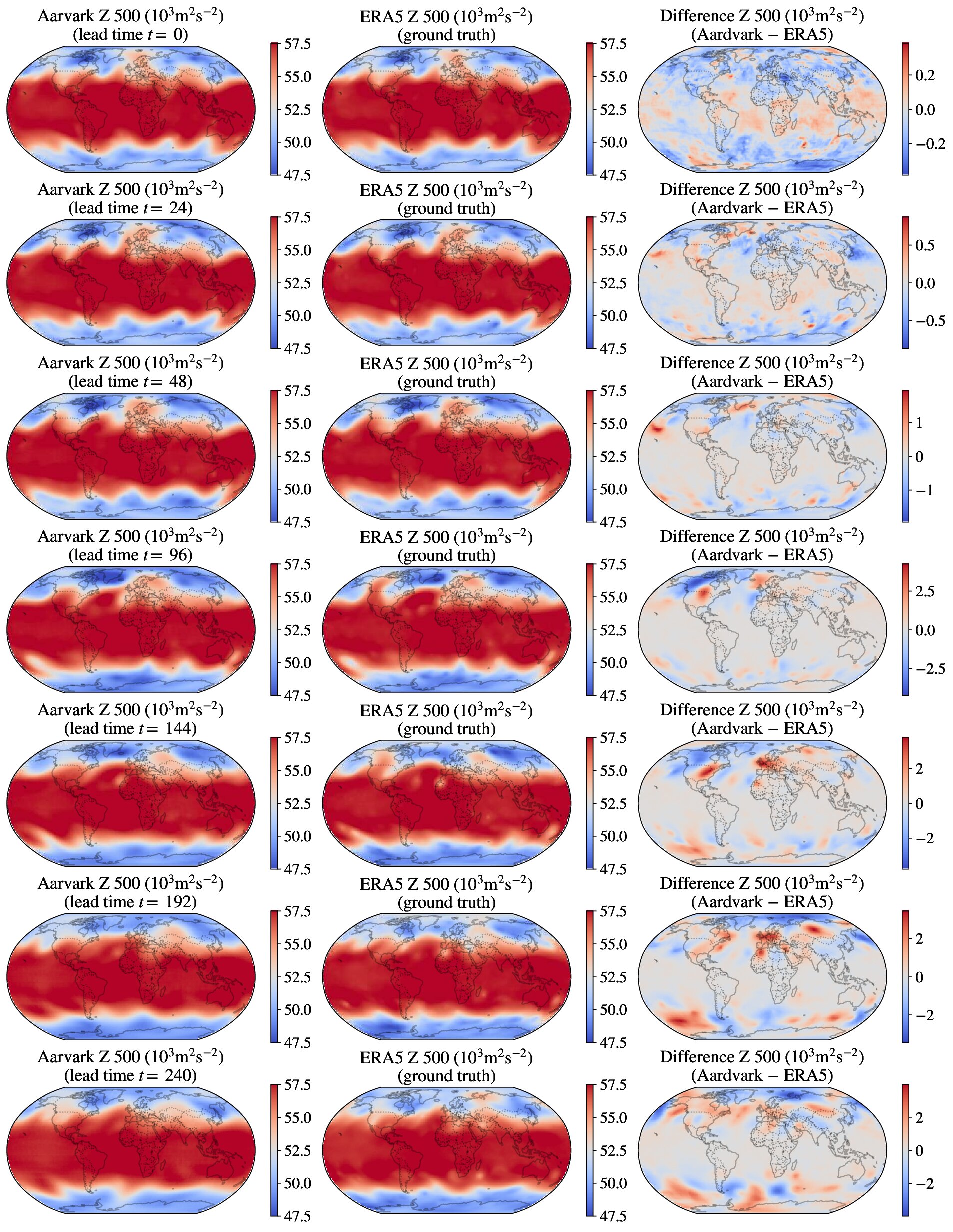}
    \caption{
        Illustration of the predictions of Aardvark.
        Note $t = 0,$ corresponds to $11^{\text{th}}$ of January 2018}
    \label{fig:app:5}
    \vspace{-5mm}
\end{figure}

\begin{figure}[h]
    \centering
    \includegraphics[width=0.90\linewidth]{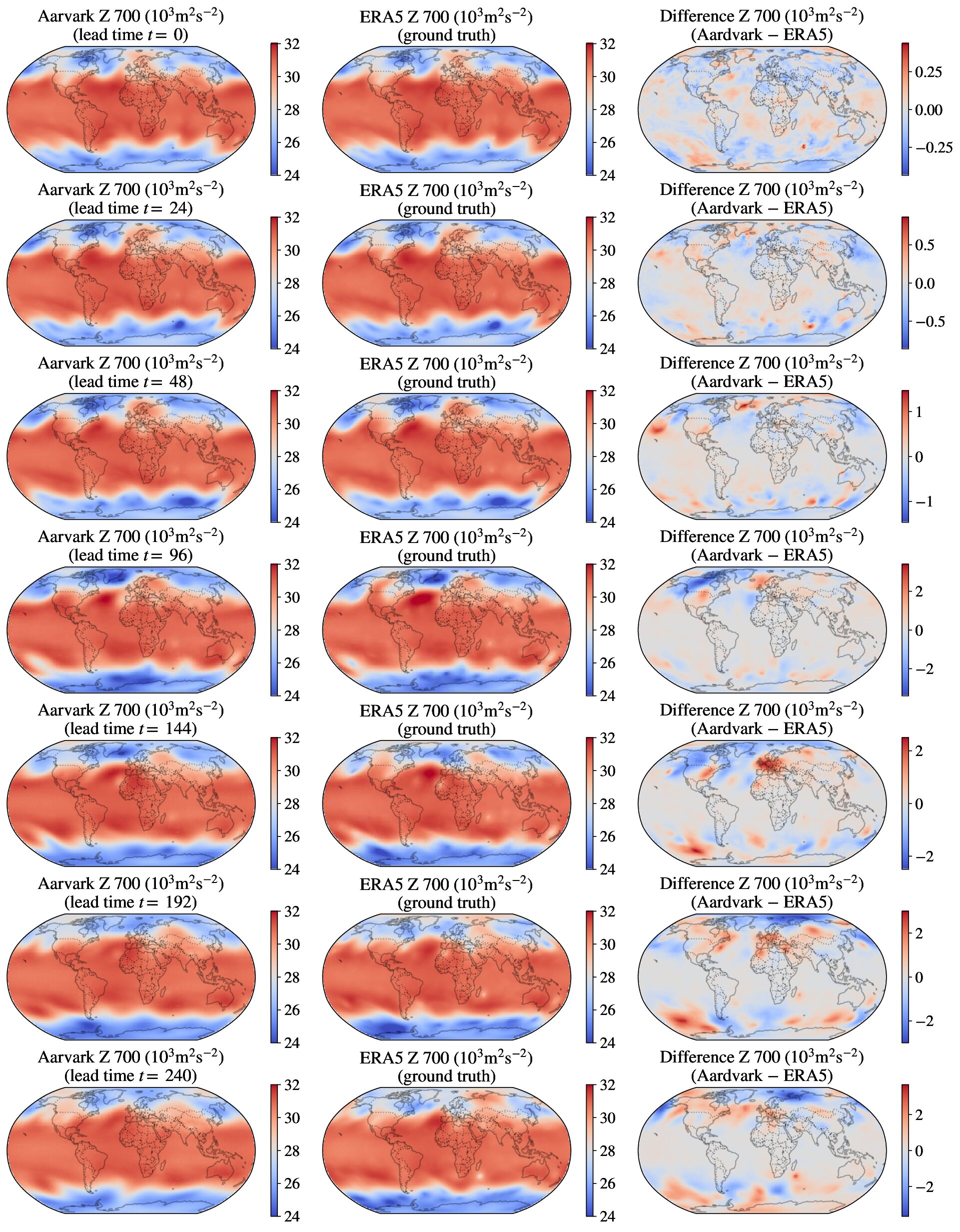}
    \caption{
        Illustration of the predictions of Aardvark.
        Note $t = 0,$ corresponds to $11^{\text{th}}$ of January 2018}
    \label{fig:app:6}
    \vspace{-5mm}
\end{figure}

\begin{figure}[h]
    \centering
    \includegraphics[width=0.90\linewidth]{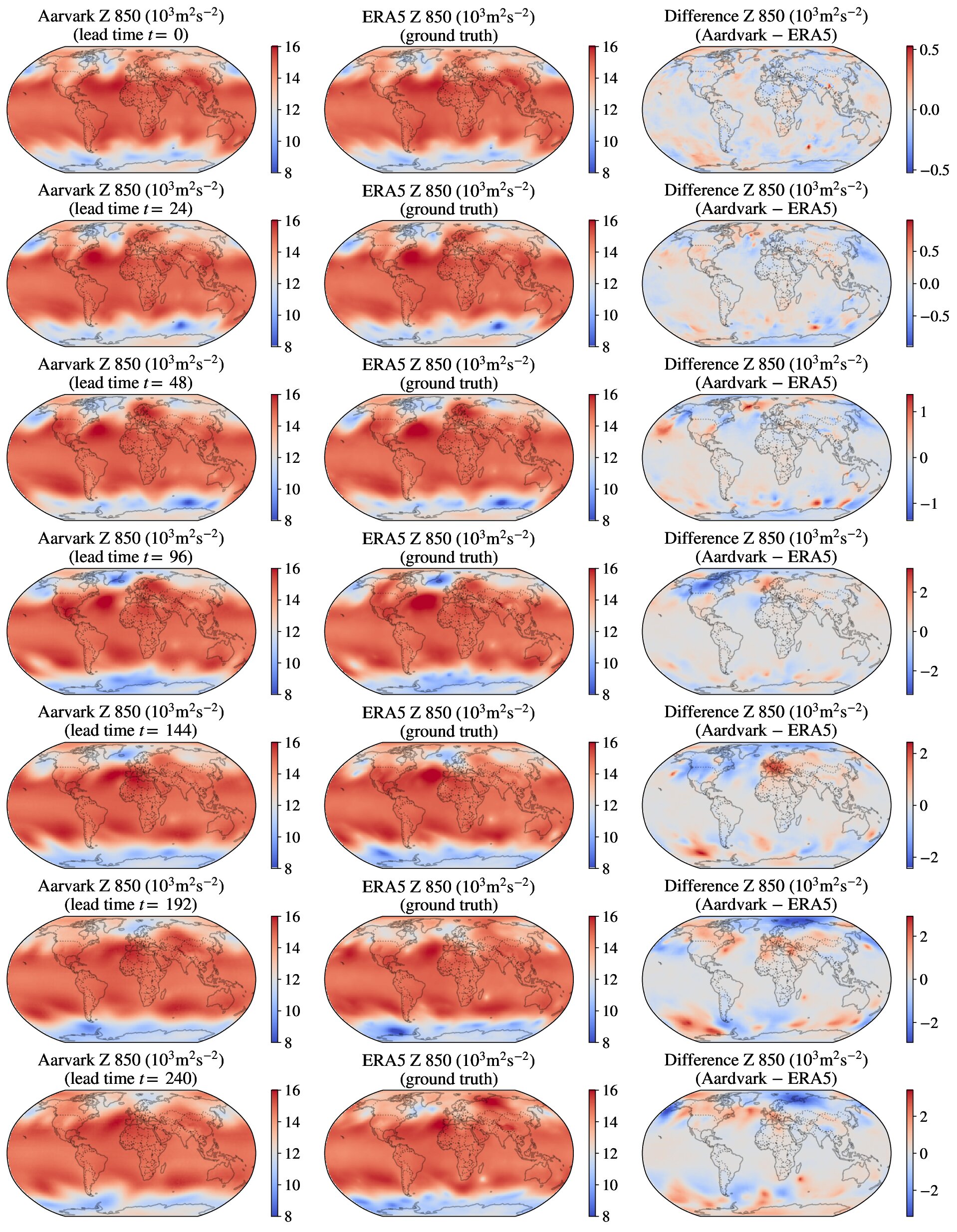}
    \caption{
        Illustration of the predictions of Aardvark.
        Note $t = 0,$ corresponds to $11^{\text{th}}$ of January 2018}
    \label{fig:app:7}
    \vspace{-5mm}
\end{figure}

\begin{figure}[h]
    \centering
    \includegraphics[width=0.90\linewidth]{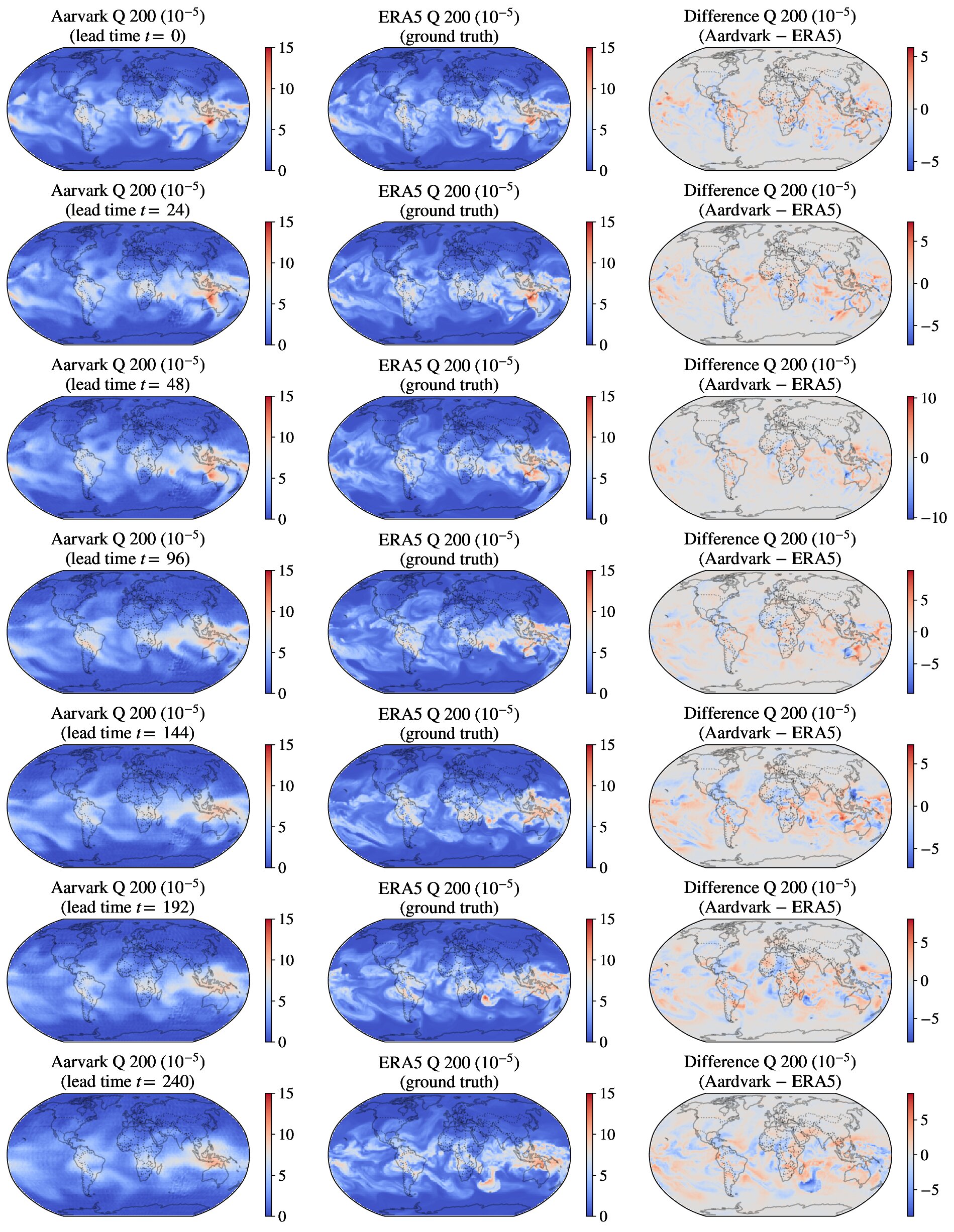}
    \caption{
        Illustration of the predictions of Aardvark.
        Note $t = 0,$ corresponds to $11^{\text{th}}$ of January 2018}
    \label{fig:app:8}
    \vspace{-5mm}
\end{figure}

\begin{figure}[h]
    \centering
    \includegraphics[width=0.90\linewidth]{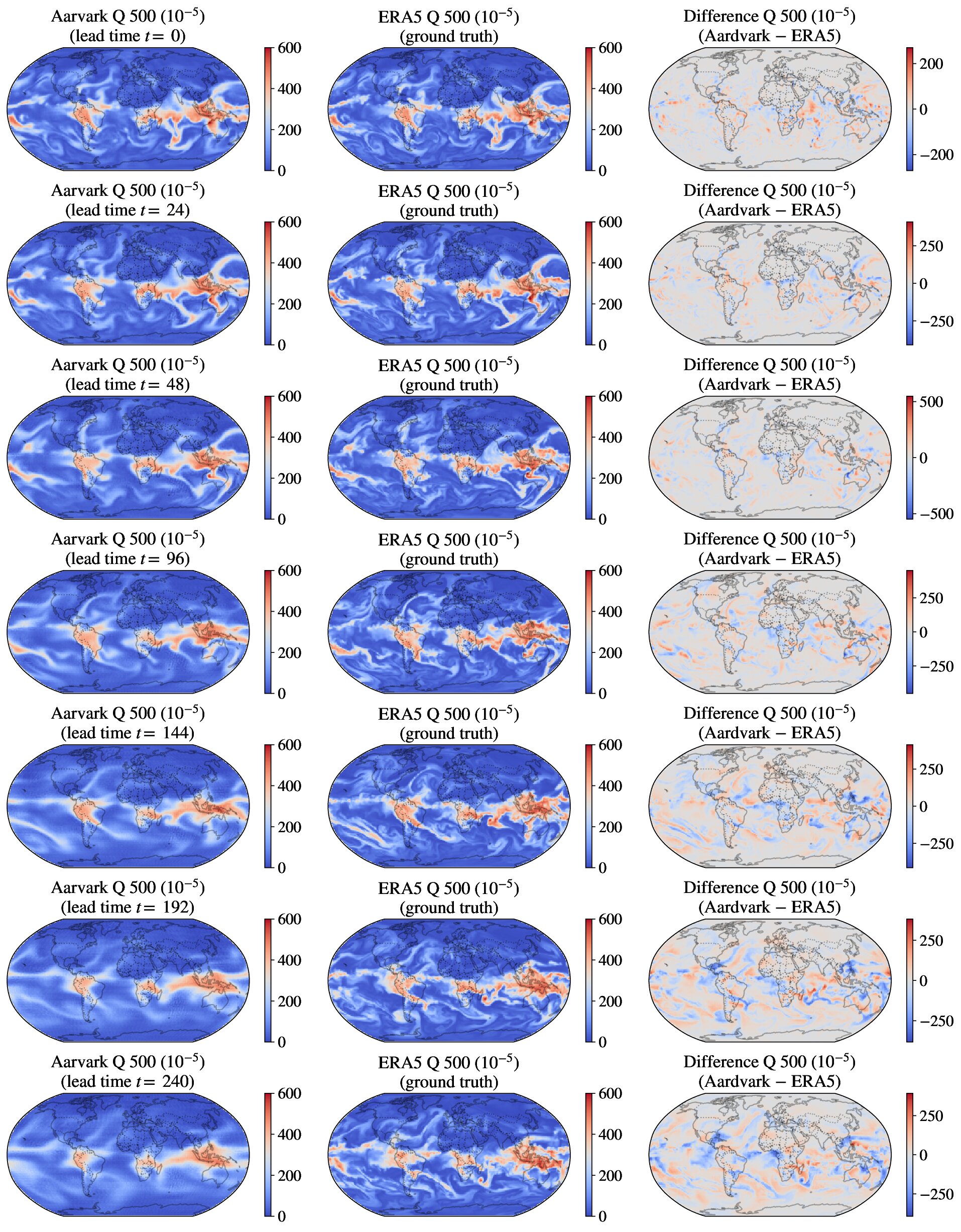}
    \caption{
        Illustration of the predictions of Aardvark.
        Note $t = 0,$ corresponds to $11^{\text{th}}$ of January 2018}
    \label{fig:app:9}
    \vspace{-5mm}
\end{figure}

\begin{figure}[h]
    \centering
    \includegraphics[width=0.90\linewidth]{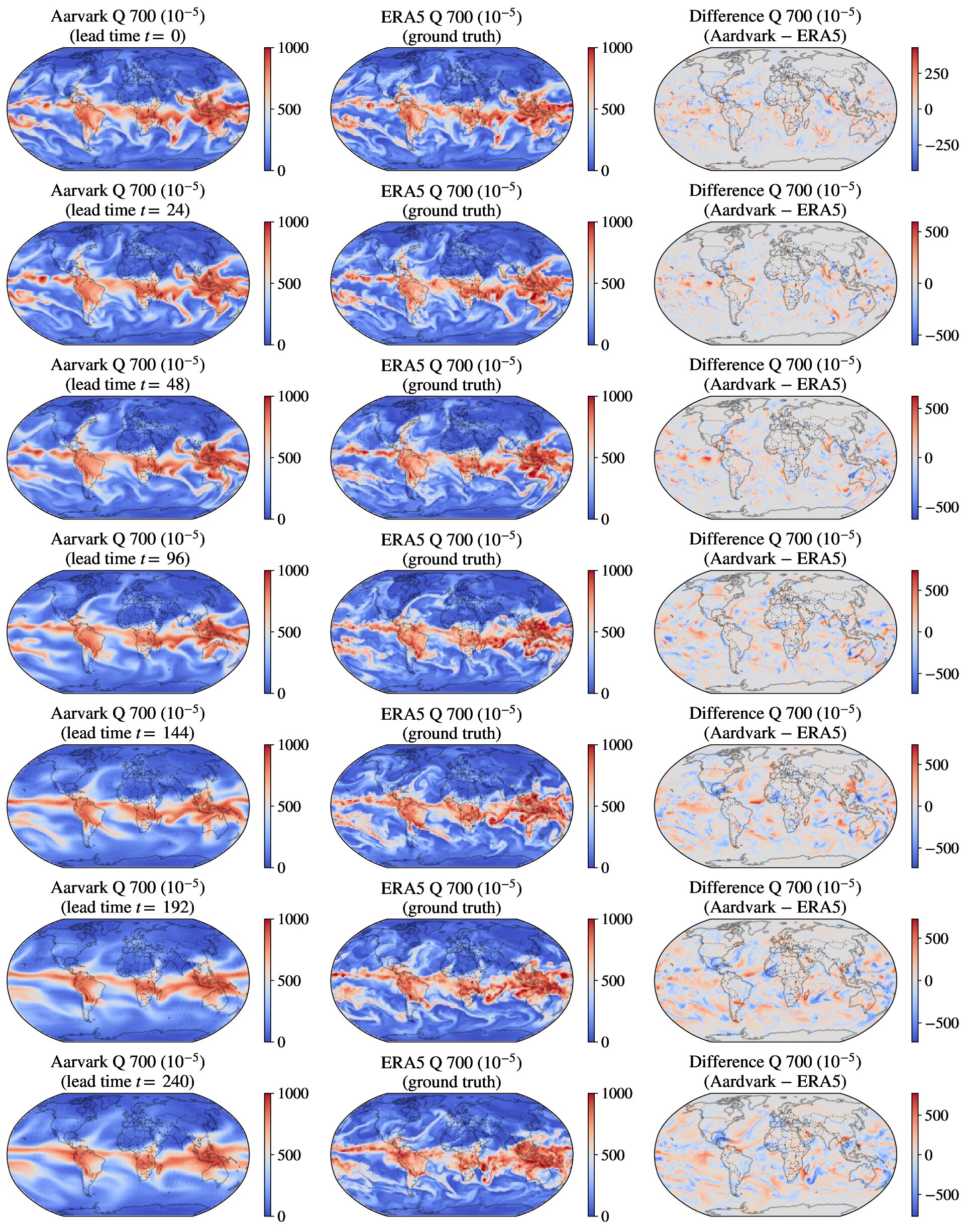}
    \caption{
        Illustration of the predictions of Aardvark.
        Note $t = 0,$ corresponds to $11^{\text{th}}$ of January 2018}
    \label{fig:app:10}
    \vspace{-5mm}
\end{figure}

\begin{figure}[h]
    \centering
    \includegraphics[width=0.90\linewidth]{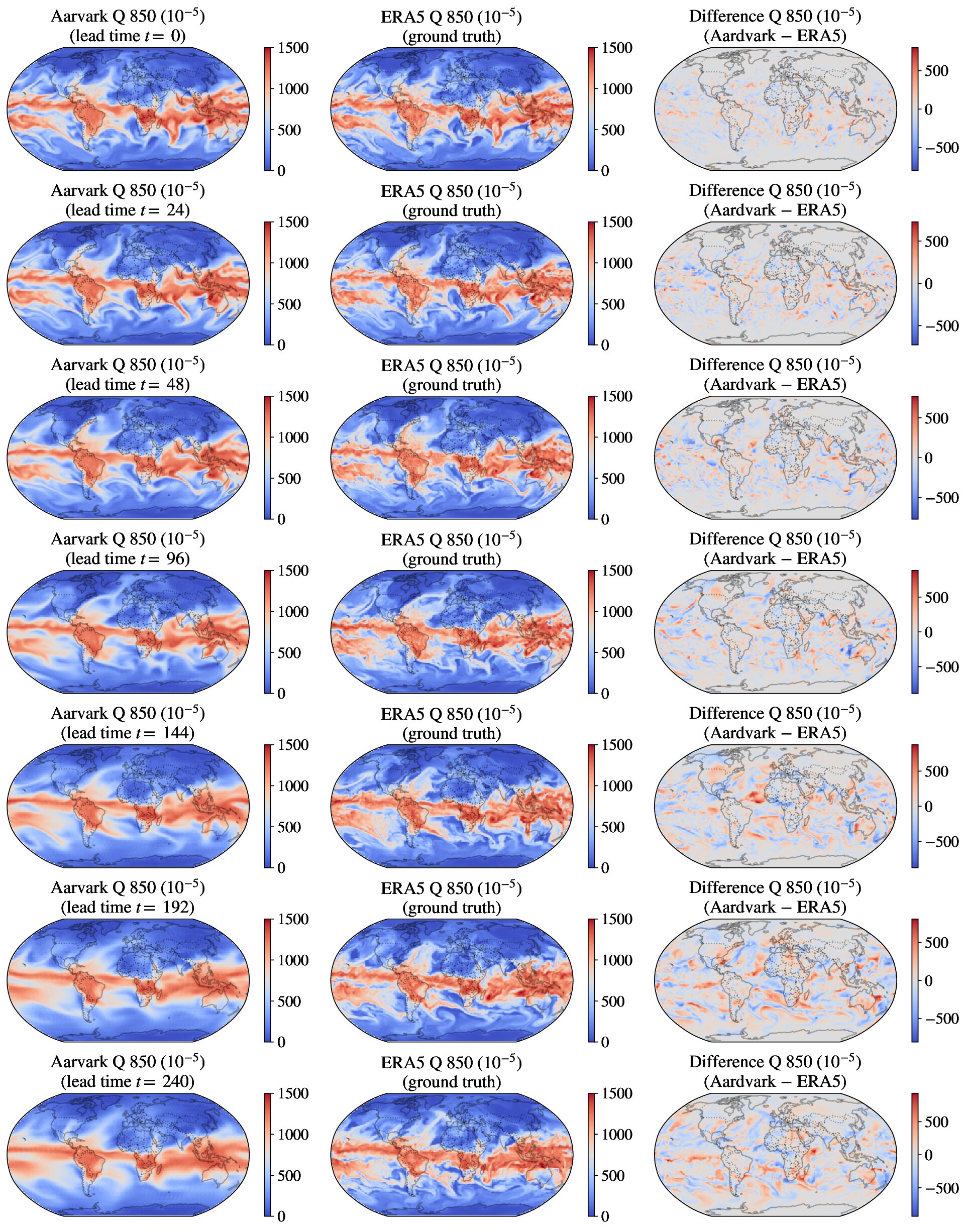}
    \caption{
        Illustration of the predictions of Aardvark.
        Note $t = 0,$ corresponds to $11^{\text{th}}$ of January 2018}
    \label{fig:app:11}
    \vspace{-5mm}
\end{figure}

\begin{figure}[h]
    \centering
    \includegraphics[width=0.90\linewidth]{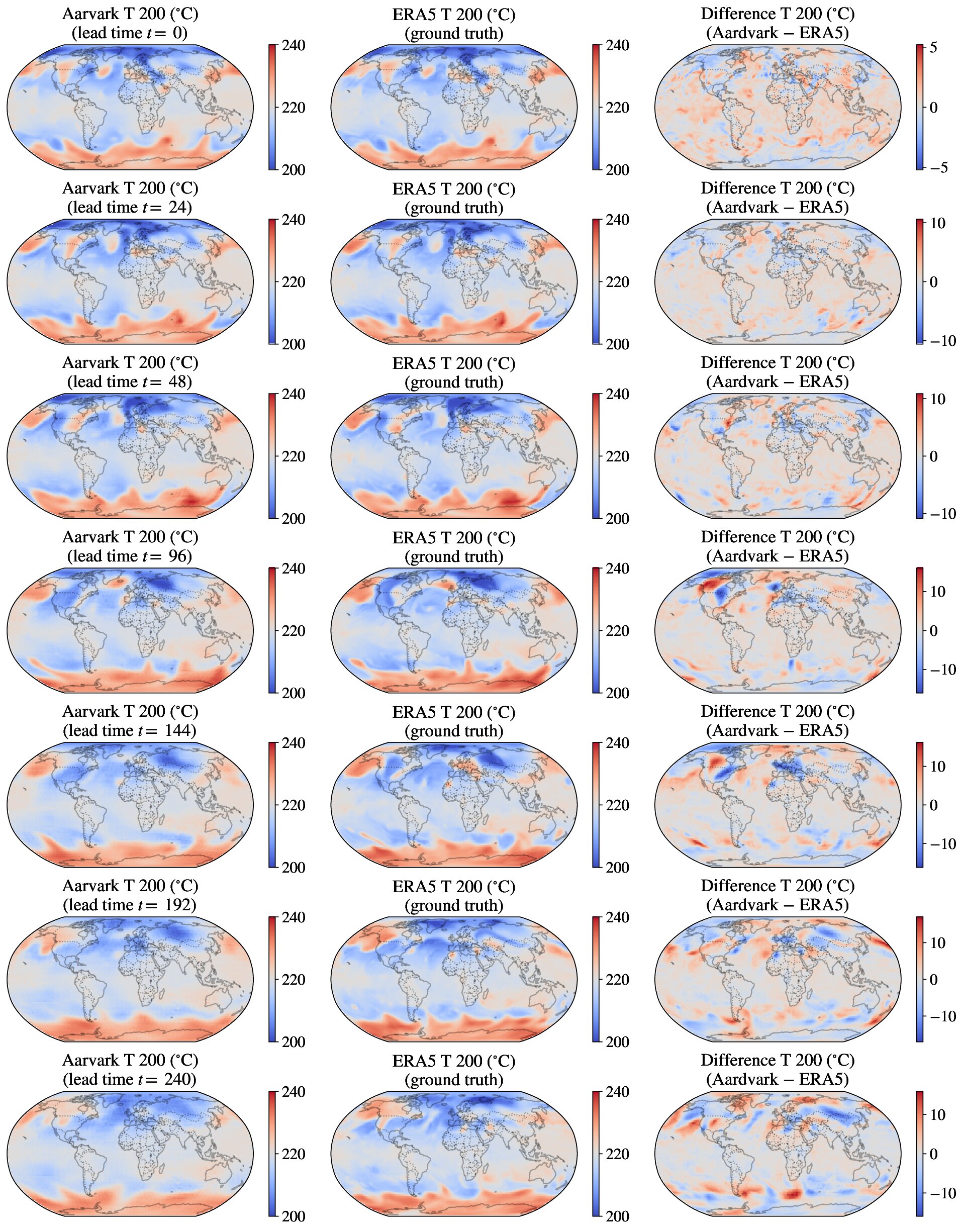}
    \caption{
        Illustration of the predictions of Aardvark.
        Note $t = 0,$ corresponds to $11^{\text{th}}$ of January 2018}
    \label{fig:app:12}
    \vspace{-5mm}
\end{figure}

\begin{figure}[h]
    \centering
    \includegraphics[width=0.90\linewidth]{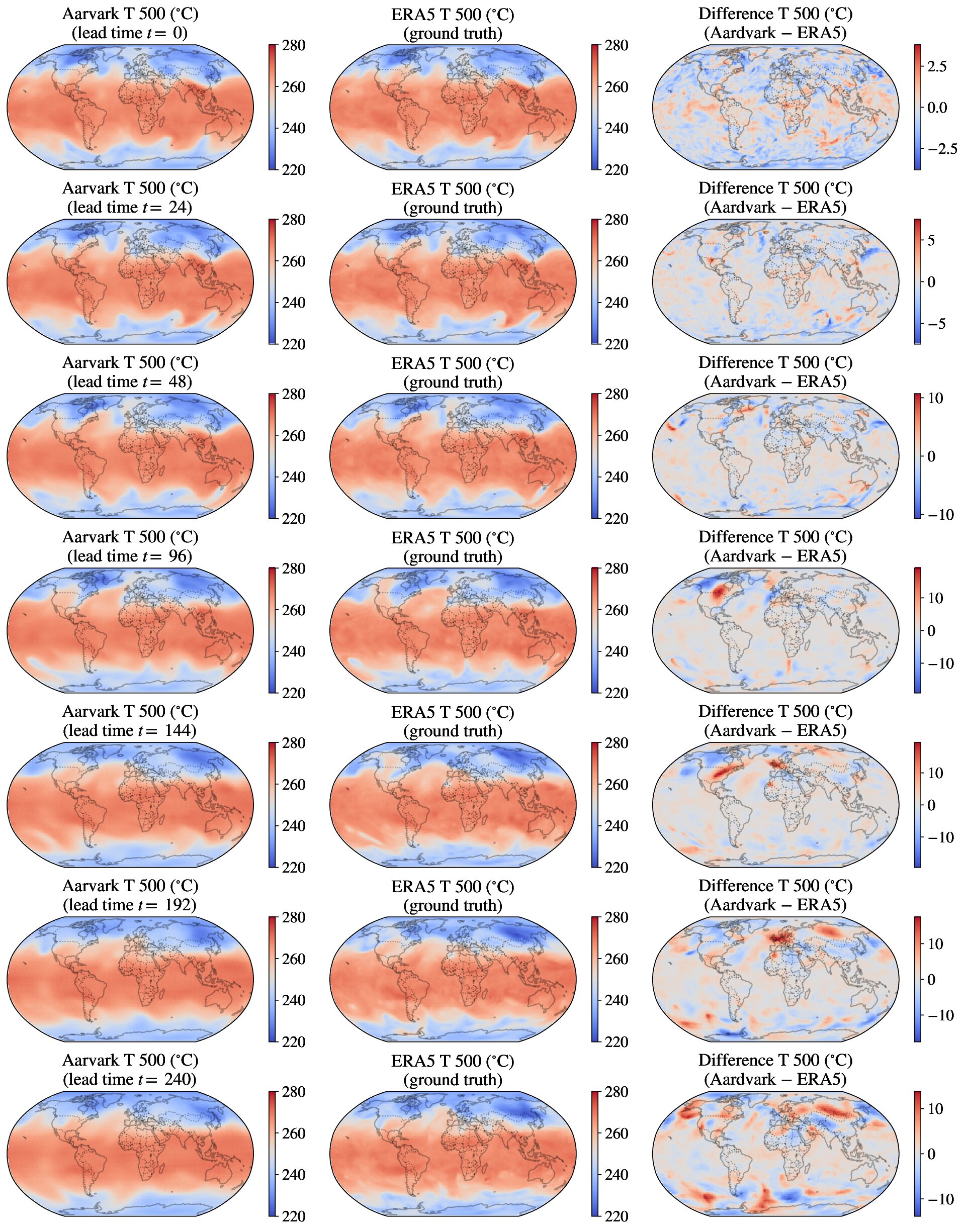}
    \caption{
        Illustration of the predictions of Aardvark.
        Note $t = 0,$ corresponds to $11^{\text{th}}$ of January 2018}
    \label{fig:app:13}
    \vspace{-5mm}
\end{figure}

\begin{figure}[h]
    \centering
    \includegraphics[width=0.90\linewidth]{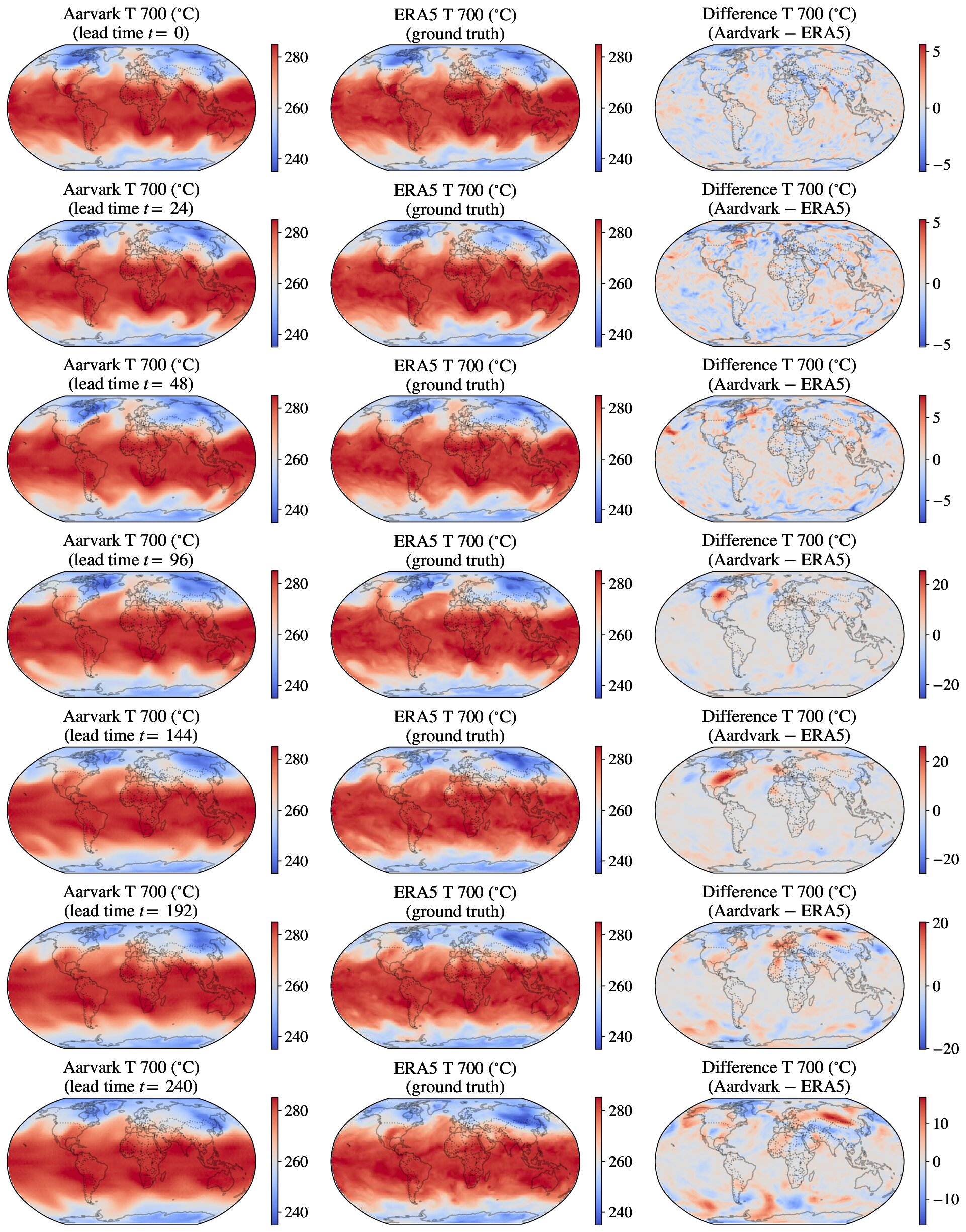}
    \caption{
        Illustration of the predictions of Aardvark.
        Note $t = 0,$ corresponds to $11^{\text{th}}$ of January 2018}
    \label{fig:app:14}
    \vspace{-5mm}
\end{figure}

\begin{figure}[h]
    \centering
    \includegraphics[width=0.90\linewidth]{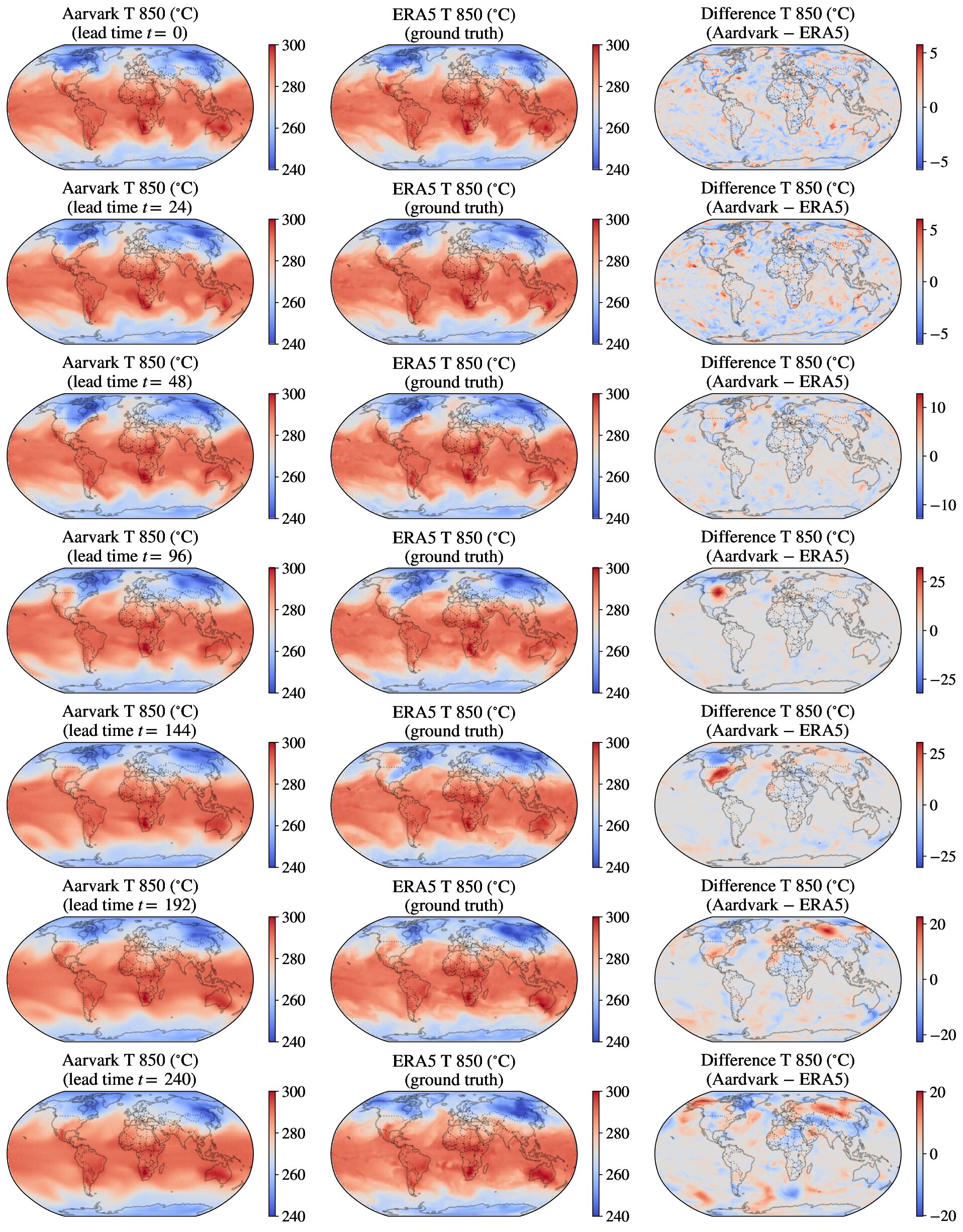}
    \caption{
        Illustration of the predictions of Aardvark.
        Note $t = 0,$ corresponds to $11^{\text{th}}$ of January 2018}
    \label{fig:app:15}
    \vspace{-5mm}
\end{figure}

\begin{figure}[h]
    \centering
    \includegraphics[width=0.90\linewidth]{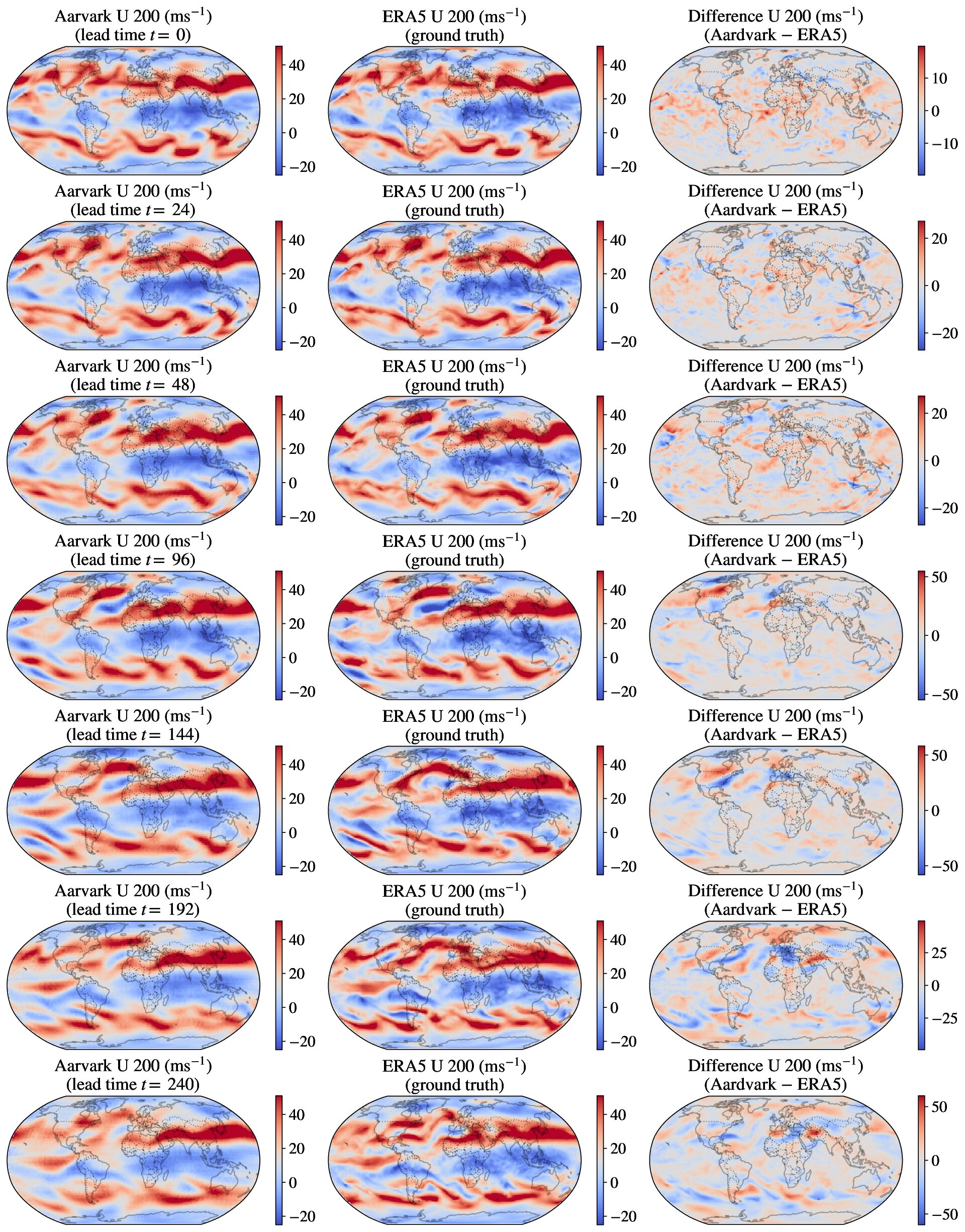}
    \caption{
        Illustration of the predictions of Aardvark.
        Note $t = 0,$ corresponds to $11^{\text{th}}$ of January 2018}
    \label{fig:app:16}
    \vspace{-5mm}
\end{figure}

\begin{figure}[h]
    \centering
    \includegraphics[width=0.90\linewidth]{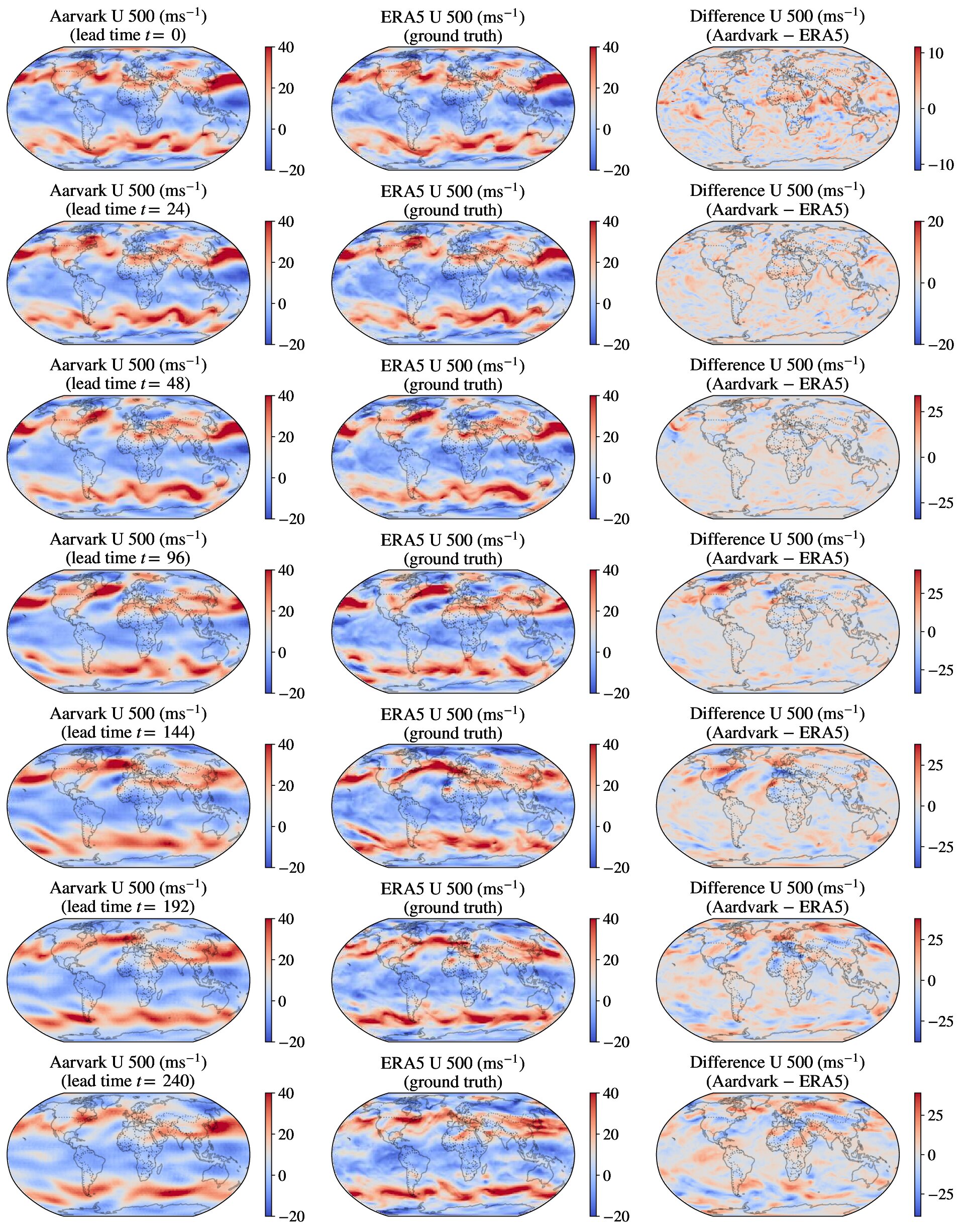}
    \caption{
        Illustration of the predictions of Aardvark.
        Note $t = 0,$ corresponds to $11^{\text{th}}$ of January 2018}
    \label{fig:app:17}
    \vspace{-5mm}
\end{figure}

\begin{figure}[h]
    \centering
    \includegraphics[width=0.90\linewidth]{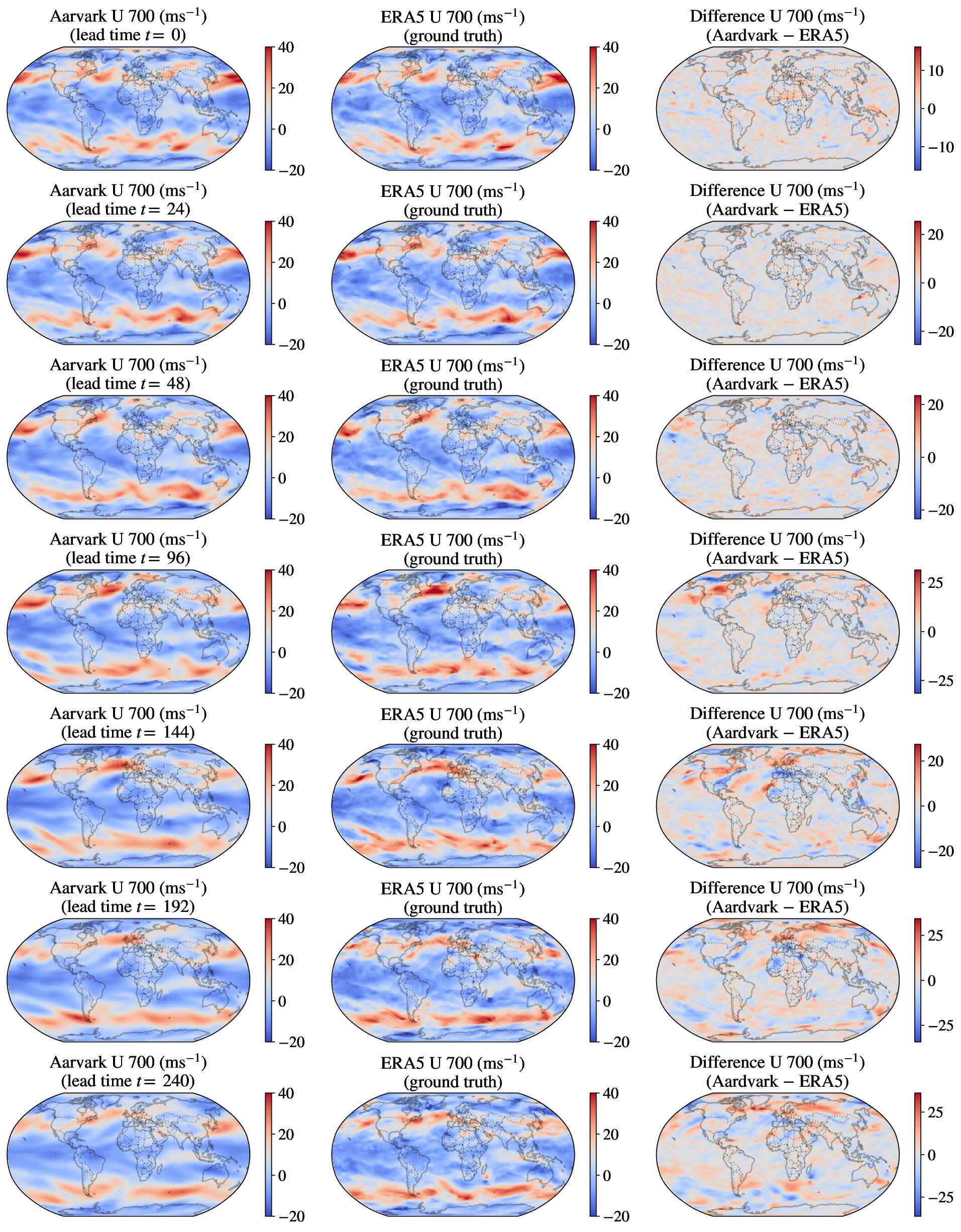}
    \caption{
        Illustration of the predictions of Aardvark.
        Note $t = 0,$ corresponds to $11^{\text{th}}$ of January 2018}
    \label{fig:app:18}
    \vspace{-5mm}
\end{figure}

\begin{figure}[h]
    \centering
    \includegraphics[width=0.90\linewidth]{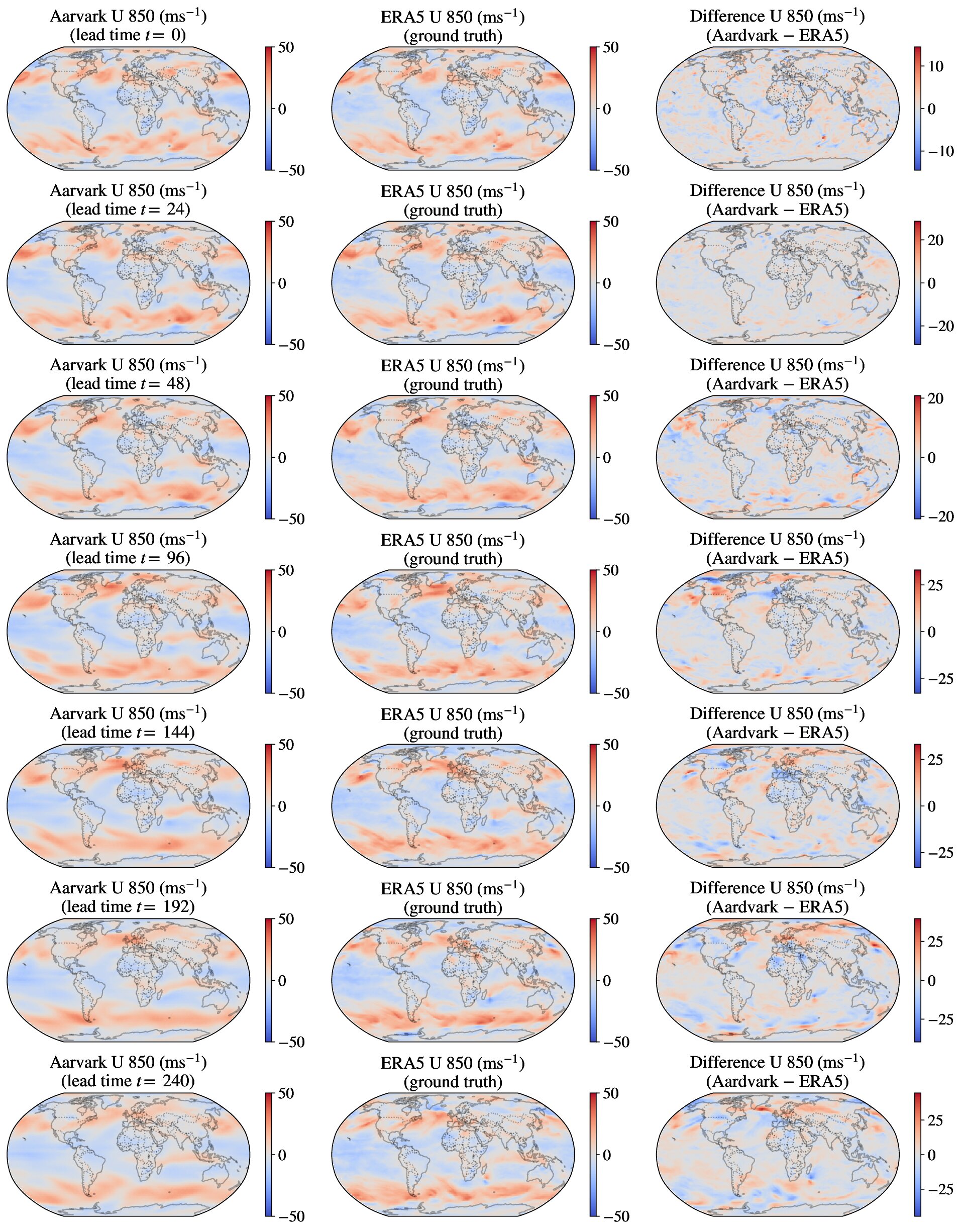}
    \caption{
        Illustration of the predictions of Aardvark.
        Note $t = 0,$ corresponds to $11^{\text{th}}$ of January 2018}
    \label{fig:app:19}
    \vspace{-5mm}
\end{figure}

\begin{figure}[h]
    \centering
    \includegraphics[width=0.90\linewidth]{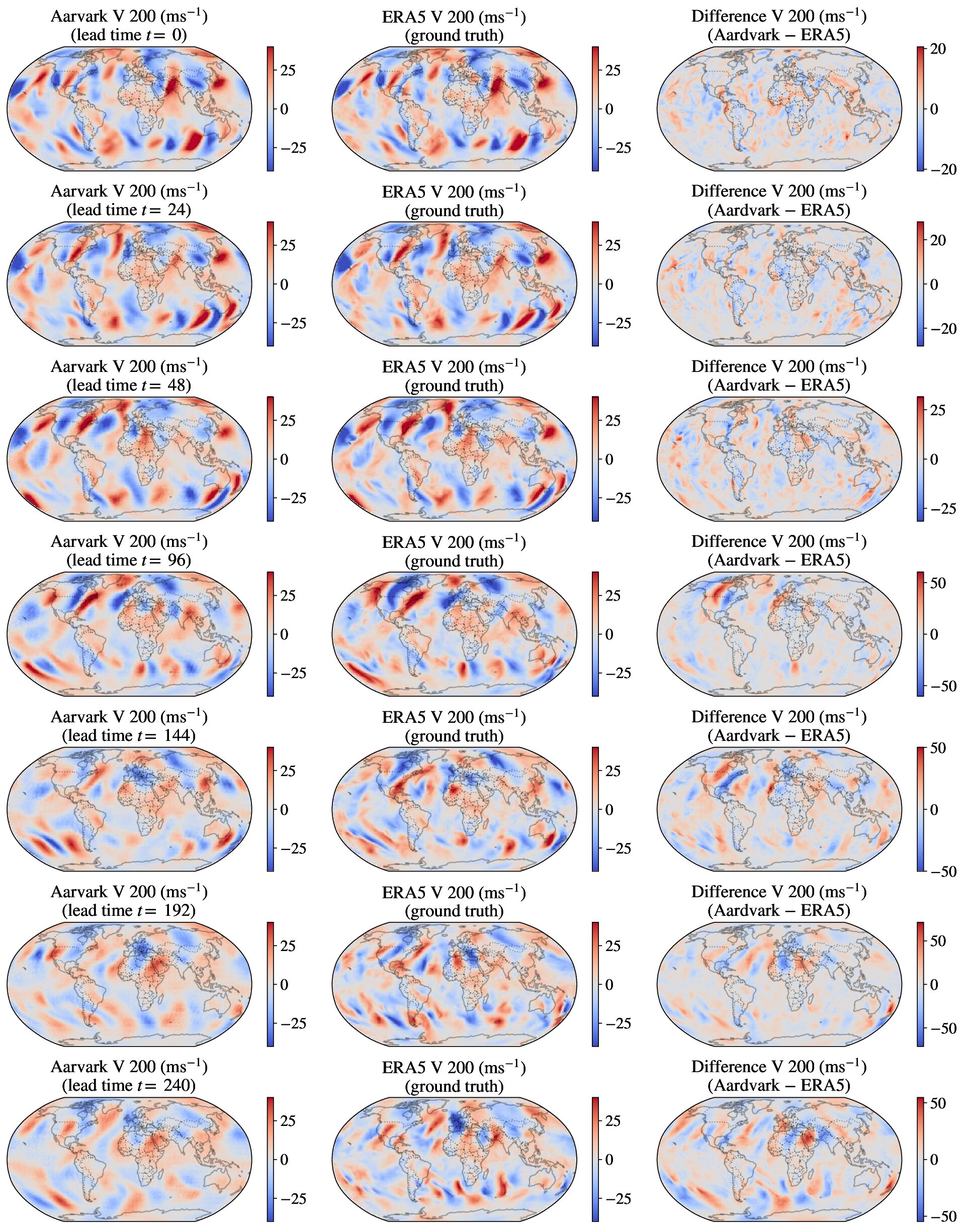}
    \caption{
        Illustration of the predictions of Aardvark.
        Note $t = 0,$ corresponds to $11^{\text{th}}$ of January 2018}
    \label{fig:app:20}
    \vspace{-5mm}
\end{figure}

\begin{figure}[h]
    \centering
    \includegraphics[width=0.90\linewidth]{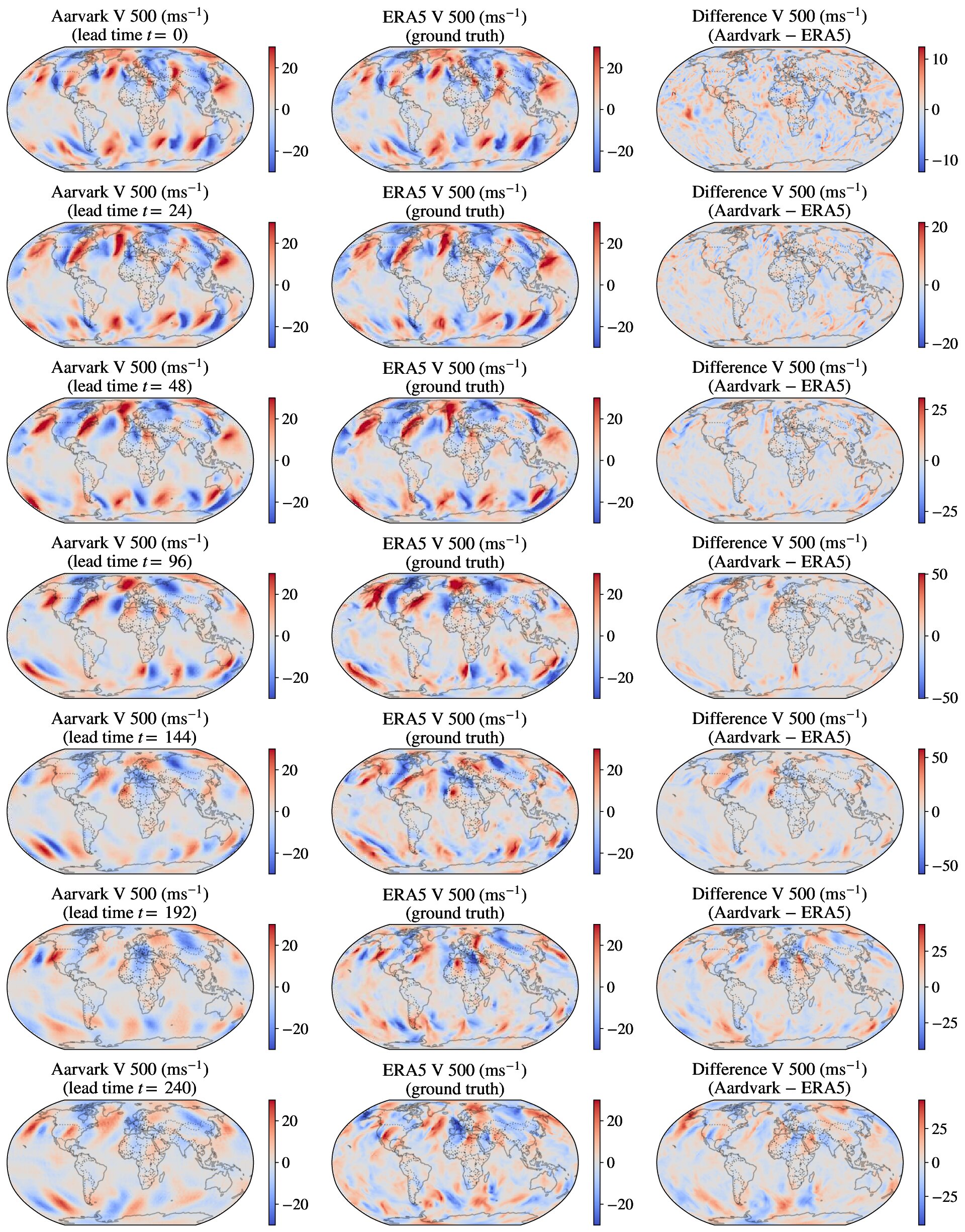}
    \caption{
        Illustration of the predictions of Aardvark.
        Note $t = 0,$ corresponds to $11^{\text{th}}$ of January 2018}
    \label{fig:app:21}
    \vspace{-5mm}
\end{figure}

\begin{figure}[h]
    \centering
    \includegraphics[width=0.90\linewidth]{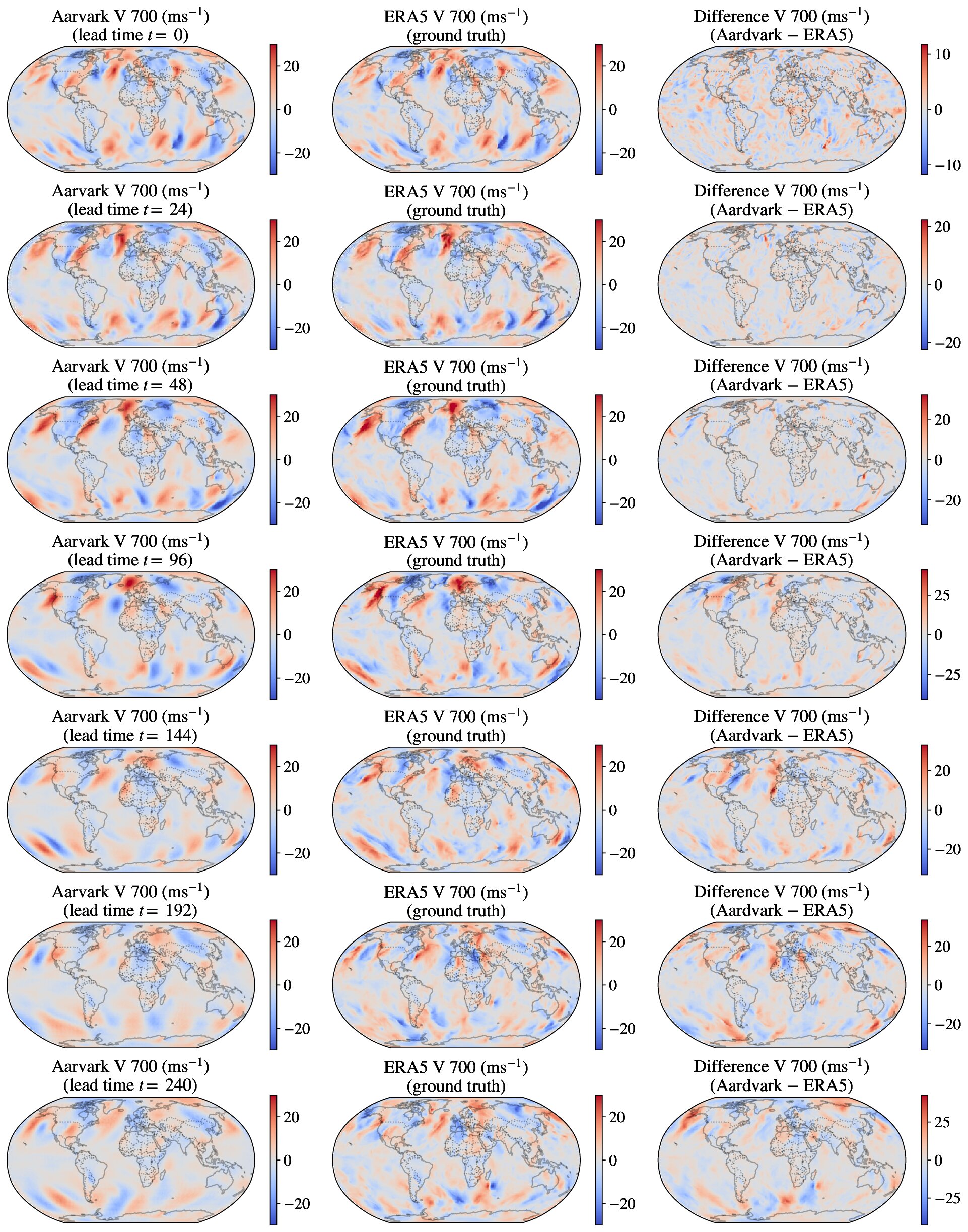}
    \caption{
        Illustration of the predictions of Aardvark.
        Note $t = 0,$ corresponds to $11^{\text{th}}$ of January 2018}
    \label{fig:app:22}
    \vspace{-5mm}
\end{figure}

\begin{figure}[h]
    \centering
    \includegraphics[width=0.90\linewidth]{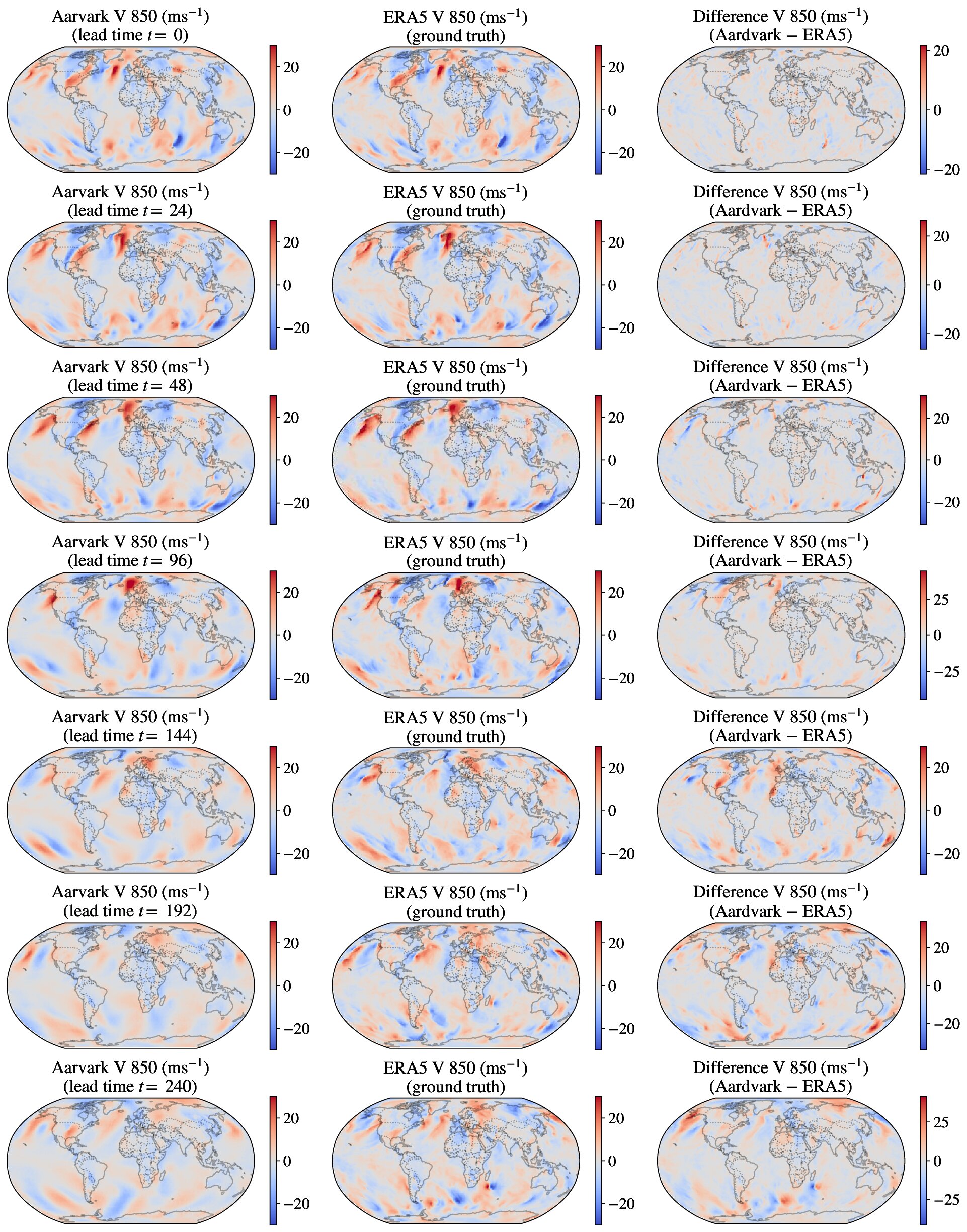}
    \caption{
        Illustration of the predictions of Aardvark.
        Note $t = 0,$ corresponds to $11^{\text{th}}$ of January 2018}
    \label{fig:app:23}
    \vspace{-5mm}
\end{figure}

\end{Backmatter}

\end{document}